\RequirePackage{fix-cm}
\documentclass[12pt,a4paper,final]{amsart}

\usepackage[export]{adjustbox}
\graphicspath{ {sections/images/} }
\usepackage{mathptmx}      
\usepackage{amssymb}
\usepackage{pifont}
\usepackage{subfiles}
\usepackage{rotating}
\usepackage{pdflscape}
\usepackage{wasysym}
\usepackage{tablefootnote}

\usepackage{graphicx}
\usepackage{multirow}
\usepackage{subcaption}
\usepackage{hyperref}
\usepackage[utf8]{inputenc}

\usepackage{bookmark}

\newcommand{\globalalphabin}{\alpha_{binary}^{gl}}
\newcommand{\alphabin}{\alpha_{binary}}
\newcommand{\cualpha}{cu\textrm{-}\alpha}
\newcommand{\Cualpha}{Cu\textrm{-}\alpha}

\newcommand{\cC}{\mathcal{C}}
\newcommand{\cS}{\mathcal{S}}

\title[Reliability in Software Engineering Qualitative Research through ICA]{Reliability in Software Engineering Qualitative Research through Inter-Coder Agreement\\\scriptsize A guide using Krippendorff's $\alpha$ \& Atlas.ti}

\author[]{\'Angel Gonz\'alez-Prieto}
\address{ETSI de Sistemas Inform\'aticos, Universidad Polit\'ecnica de Madrid. C.\ Alan Turing s/n, 28031. Madrid, Spain.}
\email{angel.gonzalez.prieto@upm.es}
\author[]{Jorge Perez}
\address{ETSI de Sistemas Inform\'aticos, Universidad Polit\'ecnica de Madrid. C.\ Alan Turing s/n, 28031. Madrid, Spain.}
\email{jorgeenrique.perez@upm.es}
\author[]{Jessica Diaz}
\address{ETSI de Sistemas Inform\'aticos, Universidad Polit\'ecnica de Madrid. C.\ Alan Turing s/n, 28031. Madrid, Spain.}
\email{yesica.diaz@upm.es}
\author[]{Daniel L\'opez-Fern\'andez}
\address{ETSI de Sistemas Inform\'aticos, Universidad Polit\'ecnica de Madrid. C.\ Alan Turing s/n, 28031. Madrid, Spain.}
\email{daniel.lopez@upm.es}

\begin{document}

\newtheorem{thm}{Theorem}[section]
\newtheorem{prop}[thm]{Proposition}
\newtheorem{lem}[thm]{Lemma}
\newtheorem{cor}[thm]{Corollary}
\newtheorem{conjecture}{Conjecture}
\newtheorem*{theorem*}{Theorem}

\theoremstyle{definition}
\newtheorem{defn}[thm]{Definition}
\newtheorem{ex}[thm]{Example}
\newtheorem{as}{Assumption}

\theoremstyle{remark}
\newtheorem{rmk}[thm]{Remark}

\date{}

\maketitle

\begin{abstract}
In recent years, the research on empirical software engineering that uses qualitative data analysis (e.g., cases studies, interview surveys, and grounded theory studies) is increasing. However, most of this research does not deep into the reliability and validity of findings, specifically in the reliability of coding in which these methodologies rely on, despite there exist a variety of statistical techniques known as Inter-Coder Agreement (ICA) for analyzing consensus in team coding. This paper aims to establish a novel theoretical framework that enables a methodological approach for conducting this validity analysis. This framework is based on a set of coefficients for measuring the degree of agreement that different coders achieve when judging a common matter. We analyze different reliability coefficients and provide detailed examples of calculation, with special attention to Krippendorff's $\alpha$ coefficients. We systematically review several variants of Krippendorff's $\alpha$ reported in the literature and provide a novel common mathematical framework in which all of them are unified through a universal $\alpha$ coefficient. Finally, this paper provides a detailed guide of the use of this theoretical framework in a large case study on DevOps culture. We explain how $\alpha$ coefficients are computed and interpreted using a widely used software tool for qualitative analysis like Atlas.ti. We expect that this work will help empirical researchers, particularly in software engineering, to improve the quality and trustworthiness of their studies.
\end{abstract}
    
\section{Introduction}
\label{sec:introduction}

In recent years, the research on empirical software engineering that uses qualitative research methods is on the rise \cite{Ghanbari,STOL:2016,wohlin2012experimentation,SALLEH:2018,Storer}. Coding plays a key role in the qualitative data analysis process of case studies, interview surveys, and grounded theory studies. Content analysis, thematic analysis and grounded theory’s coding methods (e.g. in vivo, process, initial, focused, axial, and theoretical coding) have been established as top notch procedures for conducting qualitative data analysis as they provide methods for examining and interpreting qualitative data to understand what it represents \cite{cruzes2011recommended,Saldana2012}. 

Reliability in coding is particularly crucial to identify mistakes before the codes are used in developing and testing a theory or model. In this way, it assesses the soundness and correctness of the drawn conclusions, with a view towards creating well-posed and long-lasting knowledge. Weak confidence in the data only leads to uncertainty in the subsequent analysis and generate doubts on findings and conclusions. In Krippendorff's own words: ``If the results of reliability testing are compelling, researchers may proceed with the analysis of their data. If not, doubts prevail as to what these data mean, and their analysis is hard to justify'' \cite{Krippendorff:2018}.

This problem can be addressed by means of well-established statistical techniques known as Inter-Coder Agreement (ICA) analysis. These are a collection of coefficients that measure the extend of the agreement/disagreement between several judges when they subjectively interpret a common reality. In this way, these coefficients allow researchers to establish a value of reliability of the coding that will be analyzed later to infer relations and to lead to conclusions. Coding is reliable if coders can be shown to agree on the categories assigned to units to an extent determined by the purposes of the study \cite{craggs2005evaluating}.

The ICA techniques are used in many different research context both from social sciences and engineering. For instance, these methods have been applied for evaluating the selection criteria of primary studies or data extraction in systematic literature reviews and the ratings when coding qualitative data. This has led to a variety of related terms interchangeably used, such as inter-coder, inter-rater or inter-judge agreement, according to the customary of the topic.

However, although more and more researchers apply inter-rater agreement to assess the validity of their results in systematic literature reviews and mapping studies as we analyzed in \cite{perez2020systematic}, it is a fact that few studies in software engineering analyze and test reliability and trustworthiness of coding, and thus, the validity of their findings. A systematic search in the main scientific repositories (namely, ACM Digital Library, Science Direct and Springer\footnote{The search string used is \texttt{``inter-coder agreement AND software engineering AND qualitative research''}.}) returns no more than 25 results. 
Similar results were obtained in a systematic literature review reported by Nili et al. \cite{Nili:2020} in information management research. Nevertheless, the amount of publications that test the reliability of their coding, and consequently, of their findings is notably higher in other areas, specially in health sciences, social psychology, education, and business \cite{Nili:2020}. In this paper we focus on testing reliability of qualitative data analysis, specifically reliability of coding, and in this context, judges or raters are referred to as coders.

In this paper, we propose to introduce the ICA analysis techniques in software engineering empirical research in order to enhance the reliability of coding in qualitative data analysis and the soundness of the results. For that purpose, in Section \ref{sec:background} we review some of the coefficients that, historically, have been reported in the literature to measure the ICA. We start discussing, in Section \ref{sec:background}, some general purpose statistics, like Cronbach's $\alpha$ \cite{cronbach1951coefficient}, Pearson's $r$ \cite{pearson1895vii} and Spearman's $\rho$ \cite{Spearman}, that are typically misunderstood to be suitable for ICA analysis, since they cannot measure the degree of agreement among coders but correlation, a weaker concept than agreement. Section \ref{sec:background-percent-agreement} reviews some coefficients that evaluate agreement between coders, which are not suitable for measuring reliability because they do not take into account the agreement by chance, such as the percent agreement \cite{feng2014intercoder,feng2015mistakes} and the Holsti Index \cite{holsti1969content}. We also present a third group, with representants like Scott's $\pi$ \cite{scott1955reliability} (Section \ref{sec:background-scott}), Cohen's $\kappa$ \cite{cohen1960coefficient} (Section \ref{sec:background-cohen}) and Fleiss' $\kappa$ \cite{fleiss1971measuring} (Section \ref{sec:background-fleiss}), which have been intensively used in the literature for measuring reliability, specially in social sciences. However, as pointed out by Krippendorff in \cite{hayes2007answering}, all of these coefficients suffer some kind of weakness that turns them non-optimal for measuring reliability. Finally, in Section \ref{sec:background-krip-alpha} we sketch briefly Krippendorff's proposal to overcome these flaws, the so-called Krippendorff's $\alpha$ coefficient \cite{Krippendorff:2018}, on which we will focus along of this paper.

Despite the success and wide spread of Krippendorff's $\alpha$, there exists in the literature plenty of variants of this coefficient formulated quite ad hoc for very precise and particular situations, like \cite{krippendorff1970estimating,Krippendorff:1995,yang2011coefficient,gwet2011krippendorff,de2012calculating,Krippendorff:2016,Krippendorff:2018}. The lack of uniform treatment of these measures turns their use confusing, and the co-existence of different formulations and diffuse interpretations becomes their comparison a hard task. To address this problem, Sections \ref{sec:universal-alpha} and \ref{sec:theoretical-framework} describe a novel theoretical framework that reduces the existing variants to a unique universal $\alpha$ coefficient by means of labels that play the role of meta-codes. With this idea in mind, we focus on four of the most outstanding and widely used $\alpha$ coefficients and show how their computation can be reduced to the universal $\alpha$ by means of a simple re-labelling. This framework provides new and more precise interpretations of these coefficients that will help to detect flaws in the coding process and to correct them easily on the fly. Moreover, this interpretation in terms of labels sheds light to some awkward behaviors of the $\alpha$ coefficients that are very hard to understand otherwise.

Section \ref{sec:atlas} includes a tutorial on the use and interpretation of Krippendorff's $\alpha$ coefficients for providing reliability in software engineering case studies through the software tool Atlas.ti. This tool provides support for the different tasks that take place during qualitative data analysis, as well as the calculation of the ICA measures. There exists in the market a variety of available tools with a view towards qualitative research, like NVIVO, MaxQDA, Qualcoder, Qcoder, etc., but for its simplicity and ability for computing Krippendorff's $\alpha$ coefficients, along this tutorial we will focus on Atlas.ti. The tutorial is driven by a running example based on a real case study developed by the authors about a qualitative inquiry in the DevOps domain \cite{devops}. Additionally, we highlight several peculiarities of Atlas.ti when dealing with large corpus and sparse relevant matter. 

Finally, in Section \ref{sec:conclusions} we summarize the main conclusions of this paper and we provide some guidelines on how to apply this tutorial to case studies in qualitative research. We expect that the theoretical framework, the methodology, and the subsequent tutorial introduced in this paper may help empirical researchers, particularly in software engineering, to improve the quality and soundness of their studies.

\section{Background}
\label{sec:background}

In a variety of situations, researchers have to deal with the problem of judging some data. While the observed data is objective, the perception of each researcher is deeply subjective. In many cases, the solution is to introduce several judges, typically referred to as coders, to reduce the amount of subjectivity by comparing their judgements. However, in this context, a method for measuring the degree of agreement that the coders achieve in their evaluations, i.e., the coding of the raw data, is required. Thus, researchers need to measure the reliability of coding. Only after establishing that reliability is sufficiently high, it makes sense to proceed with the analysis of the data.

It is worthy to mention that, although often used interchangeably, there is a technical distinction between the terms agreement and reliability. Inter-Coder Agreement (ICA) coefficients assess the extent to which the responses of two or more independent raters are concordant; on the other hand, inter-coder reliability evaluates the extent to which these raters consistently distinguish between different responses \cite{gisev2013interrater}. In other words, the measurable quantity is the ICA, and using this value we can infer reliability. In the same vein, we should not confuse reliability and validity. Reliability deals with the extent to which the research is deterministic and independent of the coders, and in this sense it is strongly tied to reproducibility; whereas validity deals with the truthfulness, with how the claims assert the truth. Reliability is a must for validity, but does not guarantee it. Several coders may share a common interpretation of the reality, so that we have a high level of reliability, but this interpretation might be wrong and biased, so the validity is really small.


There co-exists in the literature several statistical coefficients that have been applied for evaluating ICA, such as Cronbach's $\alpha$, Pearson's $r$ and Spearman's $\rho$. However, these coefficients cannot be confused with methods of inter-coder agreement test, as none of these three types of methods measure the degree of agreement among coders. For instance, Cronbach's $\alpha$ \cite{cronbach1951coefficient} is a statistic for interval or ratio level data that focuses on the consistency of coders when numerical judgements are required for a set of units. As stated in \cite{nili2017critical}: ``It calculates the consistency by which people judge units without any aim to consider how much they agree on the units in their judgments'', and as written in \cite{hayes2007answering} ``[it is] unsuitable to assess reliability of judgments''. 

On the other hand, correlation coefficients, as Pearson's $r$ \cite{pearson1895vii} or Spearman's rank $\rho$ \cite{Spearman}, measure the extent to which two logically separate interval variables, say $X$ and $Y$, covary in a linear relationship of the form $Y = a + bX$. They indicate the degree to which the values of one variable predict the values of the other. Agreement coefficients, in contrast, must measure the extent to which $Y = X$. High correlation means that data approximate to some regression line, whereas high agreement means that they approximate the 45-degrees line \cite{Krippendorff:2018}.

For these reasons, more advanced coefficients for measuring agreement are needed to assess reliability. In this section we analyze several proposals that have been reported in the literature for quantifying the Inter-Coder Agreement (ICA) and to infer, from this value, the reliability of the coding.

\subsection{Percent agreement}
\label{sec:background-percent-agreement}

This measure is computed as the rate between the number of times the coders agreed when classifying an item (a datum to be analyzed) with the total amount of items (multiplied by 100 if we want to express it as a percent). It has been widely used in the literature due to its simplicity and straightforward calculation \cite{feng2014intercoder,feng2015mistakes}. However, it is not a valid measure for inferring reliability in case studies that require a high degree of accuracy, since it does no take into account the agreement by chance \cite{nili2017critical} and, according to \cite{hayes2007answering}, there is no a clear interpretation for values different than $100\%$. In addition, it can be used only by two coders and only for nominal data \cite{zhao2013assumptions}.

Table \ref{tab:example-background} shows an illustrative example, which is part of a Systematic Literature Review (SLR) performed by some authors of this paper \cite{perez2020systematic}. This table shows the selection of primary studies; each item corresponds to one of these studies and each coder (denoted by $J_1$ and $J_2$) determines, according to a pre-established criterion, if the studies should be promoted to an analysis phase (Y) or not (N). From these data, the percent agreement attained is $(10/15) \cdot 100 = 66.7\%$. At a first sight, this seems to be a high value that would lead to a high reliability on the data. However, we are missing the fact that the coders may achieve agreement purely by chance. As pointed out by \cite{Krippendorff:2018}: ``Percent-agreement is often used to talk about the reliability between two observers, but it has no valid reliability interpretations, not even when it measures $100\%$. Without reference to the variance in the data and to chance, percent agreement is simply uninterpretable as a measure of reliability—regardless of its popularity in the literature''. Indeed, following sections show the values of other ICA measures such as Cohen's $\kappa$ and Krippendorff's $\alpha$, which are very low: 0.39 ($39\%$) and 0.34 ($34\%$), respectively. 

\begin{table}[!h]
    \centering
    \begin{tabular}{|c|c|c|c|c|c|c|c|c|c|c|c|c|c|c|c|}
         \hline Item: & \#01 & \#02 & \#03 & \#04 & \#05 & \#06 & \#07 & \#08
 \\\hline
         $J_1$ & N & N & N & N & N & N & Y & N \\\hline
         $J_2$ & Y & Y & N & N & N & Y & Y & Y \\\hline
    \end{tabular}
    \begin{tabular}{|c|c|c|c|c|c|c|c|c|c|c|c|c|c|c|c|}
         \hline Item: & \#09 & \#10 & \#11 & \#12 & \#13 & \#14 & \#15
 \\\hline
         $J_1$ & N & Y & Y & N & N & N & Y \\\hline
         $J_2$ & N & Y & Y & N & Y & N & Y \\\hline
    \end{tabular}
    \caption{Decision table of two coders in a SLR}
    \label{tab:example-background}
\vspace{-0.3cm}
\end{table}


It is worthy to mention that there exists a variation of the simple percent agreement called the Holsti index \cite{holsti1969content}. It allows researchers to consider that the case in matter to be analyzed is not pre-divided into items to be judged, so that each coder selects the matter that considers relevant. Nevertheless, by the same reasons that the simple percent agreement, this index is not a valid measure for analyzing ICA.

\subsection{Scott's $\pi$}
\label{sec:background-scott}

This index, introduced in \cite{scott1955reliability}, is an agreement coefficient for nominal data and two coders. The method corrects percent agreement by taking into account the agreement that can occur between the coders by chance. The index of Inter-Coder Agreement for Scott's $\pi$ coefficient is computed as
$$
    \pi = \frac{P_o - P_e}{1-P_e}.
$$
Here, $P_o$ (observed percent agreement) represents the percentage of judgments on which the two analysts agree when coding the same data independently; and $P_e$ is the percent agreement to be expected on the basis of chance. This later value can be computed as
$$
    P_e = \sum_{i = 1}^k p_i^2,
$$
where $k$ is the total number of categories and $p_i$ is the proportion of the entire sample which falls in the $i$-th category.

For instance, for our example of Table \ref{tab:example-background}, we have that $P_o = 0.667$, as computed in Section \ref{sec:background-percent-agreement}. For $P_e$, it is given by $P_e = (13/30)^2 + (17/30)^2 = 0.509$. Observe that, while the number of items is $15$, in the previous computation we divided by $30$. This is due to the fact that there are $15$ items, but \emph{$30$ pairs of evaluations}, see also Table \ref{tab:background-expected-percent-agreement}.

\begin{table}[!h]
    \centering
    \begin{tabular}{|c|c|c|c|c|}
    \hline & $J_1$ & $J_2$ & $p_i$ & $p_i^2$ \\\hline
    Y & 4 & 9 & 4/30 + 9/30 = 0.433 & 0.188 \\\hline
    N & 11 & 6 & 11/30 + 6/30  = 0.567 & 0.321 \\\hline
    Total & & & & 0.509 \\\hline
    \end{tabular}
    \caption{Expected percent agreement for the data of Table \ref{tab:example-background}}
    \label{tab:background-expected-percent-agreement}
\vspace{-0.3cm}
\end{table}

Therefore, the Scott's $\pi$ coefficient has a value of
$$
    \pi = \frac{0.667 - 0.509}{1 - 0.509} = 0.322.
$$

\subsection{Cohen's $\kappa$}\label{sec:cohen}
\label{sec:background-cohen}

Cohen's $\kappa$ coefficient measures the concordance between two judges' classifications of $m$ elements into $k$ mutually exclusive categories. Cohen defined the coefficient as ``the proportion of chance-expected disagreements which do not occur, or alternatively, it is the proportion of agreement after chance agreement is removed from consideration'' \cite{cohen1960coefficient}. The coefficient is defined as
$$
    \kappa = \frac{P_o - P_c}{1-P_c},
$$
where $P_o$ is the proportion of units for which the coders agreed (relative observed agreement among raters) and $P_c$ is the proportion of units for which agreement is expected by chance (chance-expected agreement).

In order to compute these proportions, we will use the so-called contingency matrix, as shown in Table \ref{tab:background-contingency-matrix}. This is a square matrix of order the number of categories $k$. The $(i,j)$-entry, denoted $c_{i,j}$, is the number of times that an item was assigned to the $i$-th category by coder $J_1$ and to the $j$-th category by coder $J_2$. In this way, the elements of the form $c_{i,i}$ are precisely the agreements in the evaluations.

From this contingency matrix, the observed agreement, $P_o$, and agreement by chance, $P_c$, are defined as
 is defined as
$$
    P_o = \frac{1}{m}\sum_{i = 1}^k c_{i,i}, \quad P_c = \sum_{i=1}^k p_{i},
$$
where the probability of the $i$-th category, $p_i$, is given by
$$
    p_i = \left(\frac{1}{m}\sum_{j=1}^k c_{i,j}\right)\left(\frac{1}{m}\sum_{j=1}^k c_{j,i}\right) = \frac{1}{m^2} \left(\sum_{j=1}^k c_{i,j}\right)\left(\sum_{j=1}^k c_{j,i}\right).
$$

\small
\begin{table}[!h]
    \centering
    \begin{tabular}{|c|c|c|c|c|c|}
         \hline \multicolumn{2}{|c|}{\multirow{2}{*}{}} & \multicolumn{4}{c|}{$J_1$}  \\\cline{3-6}
         \multicolumn{2}{|c|}{} & Category 1 & Category 2 & $\ldots$ & Category $k$ \\\hline
         \multirow{4}{*}{$J_2$} & Category 1 & $c_{1,1}$  & $c_{2,1}$ & $\ldots$ & $c_{k,1}$ \\\cline{2-6}
         & Category 2 & $c_{1,2}$ & $c_{2,2}$ & &  $\cdots$ \\\cline{2-6}
         & $\cdots$ & & & & \\\cline{2-6}
         & Category $k$ & $c_{1,k}$ & $\ldots$ & & $c_{k,k}$ \\\hline
    \end{tabular}
    \caption{Contingency matrix}
    \label{tab:background-contingency-matrix}
\end{table}
\vspace{-0.3cm}

\normalsize

The coefficient is $\kappa=0$ when the observed agreement is entirely due to chance agreement. Greater-than-chance agreement corresponds to a positive value of $\kappa$ and less-than-chance agreement corresponds to a negative value of $\kappa$. The maximum value of $\kappa$ is $1$, which occurs when (and only when) there is perfect agreement between the coders \cite{cohen1960coefficient}. Landis and Koch, in \cite{Landis-Koch}, proposed the following table for evaluating intermediate values (Table \ref{tab:interpretation-kappa}).

\begin{table}[!h]
    \centering
    \begin{tabular}{r|l}
        Cohen's $\kappa$ & Strength of Agreement \\\hline
         < 0.0 & Poor \\
         0.00- 0.20 & Slight \\
         0.21 - 0.40 & Fair \\
         0.41 - 0.60 & Moderate \\
         0.61 - 0.80 & Substantial \\
         0.81 - 1.00 & Almost perfect \\
    \end{tabular}
    \caption{Landis \& Koch: interpretation of the value of $\kappa$}
    \label{tab:interpretation-kappa}
\vspace{-0.3cm}
\end{table}

As an example of application of this coefficient, let us come back to our example of Table \ref{tab:example-background}. From the data, we compute the contingency matrix as shown in Table \ref{tab:contingecy-matrix-example}.

\begin{table}[h]
    \centering
    \begin{tabular}{|c|c|c|c|c|}
         \hline \multicolumn{2}{|c|}{\multirow{2}{*}{}} & \multicolumn{2}{c|}{$J_1$}  & \\\cline{3-5}
         \multicolumn{2}{|c|}{}  & Y & N & Total\\\hline
         \multirow{2}{*}{$J_2$} & Y & 4 & 5 & 9 \\\cline{2-5}
         & N & 0 & 6 & 6 \\\hline
         & Total & 4 & 11 & 15\\\hline
    \end{tabular}
    \caption{Contingency matrix for the data of Table \ref{tab:example-background}}
    \label{tab:contingecy-matrix-example}
\vspace{-0.3cm}
\end{table}

In this way, $P_o$ is given by
$$
    P_o = \frac{1}{m} \left(c_{1,1} + c_{2,2}\right) = \frac{4+6}{15} = 0.667.
$$
On the other hand, for $P_c$ we have
$$
    p_1 = \frac{1}{15^2}(4 + 0) (4 + 5) = \frac{36}{225}, \quad p_2 = \frac{1}{15^2}(5 + 6) (0 + 6) = \frac{66}{225}.
$$
Hence $P_c = 0.453$. Therefore, we have
$$
    \kappa = \frac{P_o - P_c}{1-P_c} = 0.391.
$$

It is worthy to mention that, despite of its simplicity, this coefficient has some intrinsic problems. On one hand, it is limited to nominal data and two coders. On the other hand, it is difficult to interpret the result. Under various conditions, the $\kappa$ statistic is affected by two paradoxes that return biased estimates of the statistic itself: (1) high levels of observer agreement with low $\kappa$ values; (2) lack of predictability of changes in $\kappa$ with changing marginals \cite{lantz1996behavior}. Some proposals for overcoming these paradoxes are described in \cite{feinstein1990high} and in \cite{cicchetti1990high}. According to \cite{hayes2007answering}: ``$\kappa$ is simply incommensurate with situations in which the reliability of data is the issue''.

\subsection{Fleiss' $\kappa$}\label{sec:fleiss}
\label{sec:background-fleiss}

Fleiss' $\kappa$ is a generalization of Scott's $\pi$ statistic to an arbitrary number, say $n \geq 2$, of raters $J_1, \ldots, J_n$. As above, we set $m$ to be the number of items to be coded and $k$ the number of categories (possible categorical ratings) under consideration.
It is important to note that whereas Cohen's $\kappa$ assumes the same two raters have rated a set of items, Fleiss' $\kappa$ specifically allows that, although there are a fixed number of raters, different items may be rated by different individuals \cite{fleiss1971measuring}.
Analogously to Scott's $\pi$, the coefficient is calculated via the formula
$$
    \kappa = \frac{P_o - P_e}{1 - P_e}.
$$



For this coefficient, we will no longer focus on the contingency matrix, but on the number of ratings. Hence, given a category $1 \leq i \leq k$ and an item $1 \leq \beta \leq m$, we will denote by $n_{i,\beta}$ the number of raters that assigned the $i$-th category to the $\beta$-th item.

In this case, for each item $\beta$ and for each category $i$, we can compute the corresponding proportion of observations as
$$
    p_i = \frac{1}{nm} \sum_{\beta=1}^m n_{i,\beta}, \quad \hat{p}_{\beta} = \frac{1}{n(n-1)} \sum_{i=1}^k n_{i,\beta}(n_{i,\beta}-1) = \frac{1}{n(n-1)} \sum_{i=1}^k n_{i,\beta}^2 - n.
$$
Recall that $p_i$ is very similar to the one considered in Cohen's $\kappa$ but, now, $\hat{p}_\beta$ counts the rate of pairs coder--coder that are in agreement (relative to the number of all possible coder--coder pairs)

In this way, the observed agreement, $P_o$, and the expected agreement, $P_e$, are the average of these quantities relative to the total number of possibilities, that is
$$
    P_o = \frac{1}{m} \sum_{\beta = 1}^m \hat{p}_{\beta}, \quad P_e = \sum_{i=1}^k p_i^2.
$$

As an example of application, consider again the Table \ref{tab:example-background}. Recall that, in our notation, the parameters of this example are $m = 15$ is the number of items (primary studies in our case), $n = 2$ is the number of coders and $k = 2$ is the number of nominal categories (Y and N in our example which are categories $1$ and $2$, respectively).

From these data, we form the Table \ref{tab:background-proportion-observations} with the computation of the counting values $n_{i, \beta}$. In the second and the third row of this table we indicate, for each item, the number of ratings it received for each of the possible categories (Y and N). For instance, for item $\#01$, $J_1$ voted it for the category N, while $J_2$ assigned it to the category Y and, thus, we have $n_{1,1} = n_{2,1} = 1$. On the other hand, for item $\#03$ both coders assigned it the the category N so we have $n_{2,3} = 2$ and $n_{1,3} = 0$. In the last column of the table we compute the observed percentages of each category, which give the results
$$
    p_1 = \frac{1}{2 \cdot 15}(1+1+0+0+0+1+2+1+0+2+2+0+1+0+2) = \frac{13}{30} = 0.433,
$$
$$
    p_2 = \frac{1}{2 \cdot 15}(1+1+2+2+2+1+0+1+2+0+0+2+1+2+0) = \frac{17}{30} = 0.567.
$$
Observe that, as expected, $p_1 + p_2 = 1$. Therefore, $P_e = 0.433^2 + 0.567^2 = 0.508$.

\begin{table}[!h]
    \centering
    \begin{tabular}{|c|c|c|c|c|c|c|c|c|c|c|c|c|c|c|c|c|c|}
         \hline $n_{i,\beta}$ & \#01 & \#02 & \#03 & \#04 & \#05 & \#06 & \#07 & \#08 & \#09 \\\hline
         Y &  1 & 1 & 0 & 0 & 0 & 1 & 2 & 1 & 0  \\\hline
         N & 1 & 1 & 2 & 2 & 2 & 1 & 0 & 1 & 2 \\\hline
         $\hat{p}_\beta$ & 0 & 0 & 1 & 1 & 1 & 0 & 1 & 0 & 1 \\\hline
    \end{tabular}
        \begin{tabular}{|c|c|c|c|c|c|c|c|c|c|c|c|c|c|c|c|c|c|}
         \hline $n_{i,\beta}$ & \#10 & \#11 & \#12 & \#13 & \#14 & \#15 & Total & $p_i$ \\\hline
         Y &  2 & 2 & 0 & 1 & 0 & 2 & 13 & 13/30 = 0.433 \\\hline
         N & 0 & 0 & 2 & 1 & 2 & 0 & 17 & 17/30 = 0.567 \\\hline
         $\hat{p}_\beta$ & 1 & 1 & 1 & 0 & 1 & 1 & 10 & \\\hline
    \end{tabular}
    \caption{Count of the proportion of observations for Fleiss' $\kappa$ for the example of Table \ref{tab:example-background}}
    \label{tab:background-proportion-observations}
\vspace{-0.3cm}
\end{table}
\normalsize


On the other hand, for the observed percentages per item, we have two types of results. First, if for the $\beta$-th item the two coders disagreed in their ratings, we have that $\hat{p}_\beta = \frac12(1 \cdot 0 + 1 \cdot 0) = 0$. However, if both coders agreed in their ratings (regardless if it was a Y or a N), we have that $\hat{p}_\beta = \frac12(2 \cdot 1 + 0 \cdot (-1)) = 1$. Their average is the observed agreement $P_o = \frac{1}{15}(5 \cdot 0 + 10 \cdot 1) = 0.667$. Therefore, the value of the Fleiss' $\kappa$ coefficient is
$$
\kappa = \frac{0.667 - 0.508}{1 - 0.508} = 0.322.
$$  

\subsection{Krippendorff's $\alpha$}
\label{sec:background-krip-alpha}

The last coefficient that we will consider for measuring ICA is Krippendorff's $\alpha$ coefficient. Sections \ref{sec:universal-alpha} and \ref{sec:theoretical-framework} are entirely devoted to the mathematical formulation of Krippendorff's $\alpha$ and its variants for content analysis. However, we believe that, for the convenience of the reader, it is worthy to introduce this coefficient here through a working example. The version that we will discuss here corresponds to the universal $\alpha$ coefficient introduced in \cite{krippendorff1970estimating} (see also Section \ref{sec:universal-alpha}), that only deals with simple codings as the examples above. This is sometimes called binary $\alpha$ in the literature, but we reserve this name for a more involved version (see Section \ref{sec:alphabin}).

Again, we consider the data of Table \ref{tab:example-background}, which corresponds to the simplest reliability data generated by two observers who assign one of two available values to each of a common set of units of analysis (two observers, binary data). In this context, this table is called the reliability data matrix. From this table, we construct the so-called matrix of observed coincidences, as shown in Table \ref{tab:background-observed-coincidences}. This is a square matrix of order the number of possible categories (hence, a $2 \times 2$ matrix in our case since we only deal with the categories Y and N).

The way in which this table is built is the following. First, you need to count in Table \ref{tab:example-background} the number of pairs (Y, Y). In this case, $4$ items received two Y from the coders ($\#07, \#10, \#11$ and $\#15$). However, the observed coincidences matrix counts \textit{ordered pairs} of judgements, and in the previous count $J_1$ always shows up first and $J_2$ appears second. Hence, we need to multiply this result by $2$, obtaining a total count of $8$ that is written down in the (Y, Y) entry of the observed coincidences matrix, denoted $o_{1,1}$. In the same spirit, the entry $o_{2,2}$ of the matrix corresponds to the $2 \cdot 6 = 12$ ordered pairs of responses (N, N). In addition, the anti-diagonal entries of the matrix, $o_{1,2}$ and $o_{2,1}$, correspond to responses of the form (Y, N) and (N, Y). There are $5$ items in which we got a disagreement ($\#01, \#02, \#06, \#08$ and $\#14$), so, as ordered pairs of responses, there are $5$ pairs of responses (Y, N) and $5$ pairs of responses (N, Y), which are written down in the observed coincidences matrix. Finally, the marginal data, $t_1$ and $t_2$, are the sums of the values of the rows and the columns and $t = t_1 + t_2$ is twice the number of items. Observe that, by construction, the observed coincidences matrix is symmetric.

\begin{table}[h]
    \centering
    \begin{tabular}{|c|c|c|c|}
         \hline  & Y & N & \\\hline
         Y & $o_{1,1} =8$ & $o_{1,2} =5$ & $t_1 = 13$ \\\hline
         N & $o_{1,2} =5$ & $o_{2,2} =12$ & $t_2 = 17$ \\\hline
           & $t_1 =13$ & $t_2=17$ & $t = 30$\\\hline
    \end{tabular}
    \caption{Observed coincidences matrix for the data of Table \ref{tab:example-background}}
    \label{tab:background-observed-coincidences}
    \vspace{-0.3cm}
\end{table}

In this way, the observed agreement is given by
$$
    P_o = \frac{o_{1,1}}{t} + \frac{o_{2,2}}{t} = \frac{8}{30} + \frac{12}{30} = 0.67.
$$

On the other hand, as in the previous methods, we need to compare these observed coincidences with the expected coincidences by chance. This information is collected in the so-called expected coincidences matrix, as shown in Table \ref{tab:background-expected-coincidences}. The entries of this matrix, $e_{i,j}$, measure the probability of getting an ordered response $(i,j)$ entirely by chance. In our case, the expected coincidences are given by
$$
    e_{1,1}  = \frac{t_1(t_1-1)}{t-1} = \frac{13 \cdot(13-1)}{30-1} = 5.38, \quad e_{2,2}  = \frac{t_2(t_2-1)}{t-1} = \frac{17 \cdot(17-1)}{30-1} = 9.38,
$$
$$
    e_{1,2} = e_{2,1}  = \frac{t_1t_2}{t-1} = \frac{13\cdot 17}{30-1} = 7.62.
$$

\begin{table}[h]
    \centering
    \begin{tabular}{|c|c|c|c|}
         \hline  & Y & N & \\\hline
         Y & $e_{1,1} =5.38$ & $e_{1,2} = 7.62$ & $t_1 = 13$ \\\hline
         N & $e_{1,2} =7.62$ & $e_{2,2} = 9.38$ & $t_2 = 17$ \\\hline
           & $t_1 =13$ & $t_2=17$ & $t = 30$\\\hline
    \end{tabular}
    \caption{Expected coincidences matrix for the data of Table \ref{tab:example-background}}
    \label{tab:background-expected-coincidences}
\end{table}
\vspace{-0.3cm}

Therefore, the expected agreement is 
$$
    P_e = \frac{e_{1,1}}{t} + \frac{e_{2,2}}{t} = \frac{5.35}{30}+\frac{9.38}{30} = 0.49.
$$

Thus, using the same formula for the ICA coefficient as is Section \ref{sec:background-scott}, we get that Krippendorff's $\alpha$ is given by
$$
    \alpha = \frac{P_o - P_e}{1-P_e} = \frac{0.67-0.49}{1-0.49} = 0.343.
$$

As a final remark, in the context of Krippendorff's $\alpha$, it is customary to use the equivalent formulation
$$
    \alpha = 1- \frac{D_o}{D_e},
$$
where $D_o = o_{1,2}$ is the observed disagreement and $D_e = e_{1,2}$ is the expected disagreement.

\section{The universal Krippendorff's $\alpha$ coefficient}
\label{sec:universal-alpha}

Krippendorff's $\alpha$ coefficient is one of the most widely used coefficients for measuring Inter-Coder Agreement in content analysis. As we mentioned in Section \ref{sec:background-krip-alpha}, one of the reasons is that this coefficient solves many of the flaws that Cohen's $\kappa$ and Fleiss' $\kappa$ suffer. For a more detailed exposition comparing these coefficients see \cite{hayes2007answering}, and for an historical description of this coefficient check \cite{Krippendorff:2018}.

In this section, we explain the probabilistic framework that underlies Krippendorff's $\alpha$ coefficient. For this purpose, we introduce a novel interpretation that unifies the different variants of the $\alpha$ coefficient presented in the literature (see for instance \cite{krippendorff1970estimating,Krippendorff:1995,hayes2007answering,yang2011coefficient,gwet2011krippendorff,de2012calculating,Krippendorff:2016,Krippendorff:2018}). These coefficients are usually presented as unrelated and through a kind of ad hoc formulation for each problem. This turns the use of Krippendorff's $\alpha$ for the unfamiliar researcher confusing and unmotivated. For this reason, we consider that is worthy to provide a common framework in which precise interpretations and comparisons can be conducted. Subsequently, in Section \ref{sec:theoretical-framework} we will provide descriptions of this variants in terms of this universal $\alpha$ coefficient. The present formulation is an extension of the work of the authors in \cite{devops} towards a uniform formulation.

Suppose that we are dealing with $n \geq 2$ different judges, also referred to as coders, denoted by $J_1, \ldots, J_n$; as well as with a collection of $m \geq 1$ items to be judged, also referred to as quotations, denoted $I_1, \ldots, I_m$. We fix a set of $k \geq 1$ admissible `meta-codes', called labels, say $\Lambda = \left\{ l_1,\ldots, l_k\right\}$.

The task of each of the coders $J_\alpha$ is to assign, to each item $I_\beta$, a collection (maybe empty) of labels from $\Lambda$. Hence, as byproduct of the evaluation process, we get a set $\Omega = \left\{\omega_{\alpha, \beta}\right\}$, for $1 \leq \alpha \leq n$ and $1 \leq \beta \leq m$, where $\omega_{\alpha, \beta} \subseteq \Lambda$ is the set of labels that the coder $J_\alpha$ assigned to the item $I_\beta$. Recall that $\omega_{\alpha, \beta}$ is not a multiset, so every label appears in $\omega_{\alpha, \beta}$ at most once. Moreover, notice that multi-evaluations are now allowed, that is, a coder may associate more than a label to an item. This translates to the fact that $\omega_{\alpha, \beta}$ may be empty (meaning that $J_\alpha$ did not assign any label to $I_\beta$), it may have a single element (meaning that $J_\alpha$ assigned only one label) or it may have more than an element (meaning that $J_\alpha$ chose several labels for $I_\beta$).

From the collection of responses $\Omega$, we can count the number of observed pairs of responses. For that, fix $1 \leq i,j \leq k$ and set
$$
    o_{i,j} = \left|\left\{(\omega_{\alpha, \beta}, \omega_{\alpha', \beta}) \in \Omega \times \Omega \,\left|\, \alpha'\neq \alpha, l_i \in \omega_{\alpha, \beta} \textrm{ and } l_j \in \omega_{\alpha', \beta} \right.\right\}\right|.
$$

In other words, $o_{i,j}$ counts the number of (ordered) pairs of responses of the form $(\omega_{\alpha, \beta}, \omega_{\alpha', \beta}) \in \Omega \times \Omega$ that two different coders $J_\alpha$ and $J_{\alpha'}$ gave to the same item $I_\beta$ and such that $J_\alpha$ included $l_i$ in his response and $J_{\alpha'}$ included $l_j$ in his response. In the notation of Section \ref{sec:cohen}, in the case that $n = 2$ (two coders) we have that $o_{i,j} = c_{i,j} + c_{j,i}$.

\begin{rmk}\label{ref:not-voted-item}
Suppose that there exists an item $I_\beta$ that was judged by a single coder, say $J_\alpha$. The other coders, $J_{\alpha'}$ for $\alpha' \neq \alpha$, did not vote it, so $\omega_{\alpha', \beta} = \emptyset$. Then, this item $I_\beta$ makes no contribution to the calculation of $o_{i,j}$ since there is no other judgement to which $\omega_{\alpha, \beta}$ can be paired. Hence, from the point of view of Krippendorff's $\alpha$, $I_\beta$ is not taken into account. This causes some strange behaviours in the coefficients of Section \ref{sec:theoretical-framework} that may seem counterintuitive.
\end{rmk}

From these counts, we construct the matrix of observed coincidences as $M_o = \left(o_{i,j}\right)_{i,j=1}^k$. By its very construction, $M_o$ is a symmetric matrix. From this matrix, we set $t_k = \sum_{j=1}^k o_{k, j}$, which is (twice) the total number of times to which the label $l_k \in \Lambda$ was assigned by any coder. Observe that $t = \sum_{k=1}^k t_k$ is the total number of judgments. In the case that each coder evaluates each item with a single non-empty label, we have $t = nm$.

On the other hand, we can construct the matrix of expected coincidences, $M_e = \left(e_{i,j}\right)_{i,j = 1}^k$, where
$$
    e_{i,j} = \left\{\begin{array}{cc}
        \frac{t_i}{t}\frac{t_j}{t-1}t = \frac{t_it_j}{t-1} & \textrm{if }i\neq j, \\
        & \\
         \frac{t_i}{t}\frac{t_i-1}{t-1}t = \frac{t_i(t_i-1)}{t-1} & \textrm{if }i = j.
    \end{array}\right.
$$
The value of $e_{i,j}$ might be though as the average number of times that we expect to find a pair $(l_i, l_j)$, when the frequency of the label $l_i$ is estimated from the sample as $t_i/t$. It is analogous to the value of the proportion $\hat{p}_\beta$ in Section \ref{sec:fleiss}. Again, $M_e$ is a symmetric matrix.

Finally, let us fix a pseudo-metric $\delta: \Lambda \times \Lambda \to [0, \infty) \subseteq \mathbf{R}$, i.e.\ a symmetric function satisfying the triangle inequality and with $\delta(l_i,l_i)=0$ for any $l_i \in \Lambda$ (recall that this is only a pseudo-metric since different labels at distance zero are allowed). This metric is given by the semantic of the analyzed problem and, thus, it is part of the data used for quantifying the agreement. The value $\delta(l_i, l_j)$ should be seen as a measure of how similar the labels $l_i$ and $l_j$ are.
Some common choices of this metric are the following.
\begin{itemize}
    \item The discrete metric. It is given by $\delta(l_i, l_j) = 0$ if $i = j$ and $\delta(l_i, l_j) = 1$ otherwise. The discrete metric means that all the labels are equally separated and is the one that will be used along this paper.
    \item The euclidean distance. If the labels $l_i$ are vectors (for instance, if $l_i$ is a collection of real values measuring some features), we can take $\delta(l_i, l_j) = ||l_i - l_j||$, the norm of the difference vector $l_i - l_j$. In the particular case that $l_i$ are real values, this metric reduces to the absolute distance.
    \item The angular metric. If the labels $l_i$ take values in an angular measure (for instance, the deviation in some experiment), we can take $\delta(l_i, l_j) = \sin^2(l_i - l_j)$. Observe that $\delta(l_i, l_j) = 0$ if $l_i$ and $l_j$ are opposed, showing the fact that $\delta$ is only a pseudo-metric.
\end{itemize}
For subtler metrics that may be used for extracting more semantic information from the data, see \cite{Krippendorff:2016}.

From these computations, we define the observed disagreement, $D_o$, and the expected disagreement, $D_e$, as
\begin{equation}\label{eq:disagreement}
    D_o = \sum_{i=1}^k\sum_{j=1}^k o_{i,j} \delta(l_i, l_j), \hspace{0.5cm} D_e = \sum_{i=1}^k\sum_{j=1}^k e_{i,j}\delta(l_i, l_j).
\end{equation}
These quantities measure the degree of disagreement that is observed from $\Omega$ and the degree of disagreement that might be expected by judging randomly (i.e.\ by chance), respectively.

\begin{rmk}
In the case of taking $\delta$ as the discrete metric, we have another interpretation of the disagreement. Observe that, in this case, since $\delta(l_i, l_i)=0$ we can write the disagreements as
$$
    D_o = \sum_{i\neq j} o_{i,j} = t - \sum_{i=1}^k o_{i,i}, \hspace{0.5cm} D_e = \sum_{i\neq j} e_{i,j} = t - \sum_{i=1}^k e_{i,i}.
$$
The quantity $P_o = \sum_{i=1}^k o_{i,i}$ (resp.\ $P_e = \sum_{i=1}^k e_{i,i}$) can be understood as the observed (resp.\ expected) agreement between the coders. In the same vein, $t = \sum_{i,j=1}^k o_{i,j} = \sum_{i,j=1}^k e_{i,j}$ may be seen as the maximum achievable agreement. Hence, in this context, the disagreement $D_o$ (resp.\ $D_e$) is actually the difference between the maximum possible agreement and the observed (resp.\ expected) agreement.
\end{rmk}

From these data, Krippendorff's $\alpha$ coefficient is defined as
$$
    \alpha = \alpha(\Omega) = 1 - \frac{D_o}{D_e}.
$$
From this formula, observe the following limiting values:
\begin{itemize}
    \item $\alpha = 1$ is equivalent to $D_o = 0$ or, in other words, it means that there exists perfect agreement in the judgements among the coders.
    \item $\alpha = 0$ is equivalent to $D_o = D_e$, which means that the agreement observed between the judgements is entirely due to chance.
\end{itemize} 
In this way, Krippendorff's $\alpha$ can be interpreted as a measure of the degree of agreement that is achieved out of the chance. The bigger the $\alpha$ is, the better agreement is observed. A common rule-of-thumb in the literature \cite{Krippendorff:2018} is that $\alpha \geq 0.667$ is the minimal threshold required for drawing conclusions from the data. For $\alpha \geq 0.80$, we can consider that there exists statistical evidence of reliability in the evaluations. Apart from these considerations, there are doubts in the community that more partitioned interpretations, like the one of Landis $\&$ Koch of Table \ref{tab:interpretation-kappa}, are valid in this context (see \cite{Krippendorff:2018}).

\begin{rmk}
Observe that $\alpha < 0$ may only be achieved if $D_o > D_e$, which means that there is even more disagreement than the one that could be expected by chance. This implies that the coders are, consistently, issuing different judgements for the same items. Thus, it evidences that there exists an agreement between the coders to not agree, that is, to fake the evaluations. On the other hand, as long as the metric $\delta$ is non-negative, $D_o \geq 0$ and, thus, $\alpha \leq 1$.
\end{rmk}

\begin{rmk}\label{rmk:change-judges}
The observed and expected coincidences, $o_{i,j}$ and $e_{i,j}$, do not depend on the particular coders under consideration. All the coders are mathematically treated on an equal footing. In particular, this allows us to split the matter under analysis into several batches that will be coded by different teams of coders. Even more, each of these teams may have a different number of coders. In this scenario, the previous formulation of the $\alpha$ coefficient still works, and the resulting coefficient measures a kind of weighted average of the agreement of each batch.

This observation may seem particularly attractive for researches, since it allows them to parallelize the coding process with several teams of coders. However, in practice this parallelization is discouraged. As we will see in Section \ref{sec:atlas}, after each round of coding, the codebook designer gives feedback to the coders aiming to precise definitions and to bound better the limits of application of the codes. This typically leads to a remarkable improvement of the observed agreement. Nevertheless, if we are working with several coding teams, their interpretation of the codebook may be different and the given feedback might be opposed. In this way, the improvement of the ICA stagnates and subsequent refinements of the codebook do not lead to better results.
\end{rmk}

\section{Theoretical framework: Semantic domains and variants of the $\alpha$ coefficient}
\label{sec:theoretical-framework}

The setting described in Sections \ref{sec:background} and \ref{sec:universal-alpha} only addresses the problem of evaluating the agreement when the judges use labels to code data. This might be too restrictive for the purposes of qualitative data analysis, where the codes typically form a two-layer structure: semantic domains that encompass broad interrelated concepts and codes within the domains that capture subtler details. This is, for instance, the setting considered by qualitative analysis tools like the Atlas.ti software \cite{Atlas:2019}.

In this section, we describe a more general framework that enables evaluating reliability in this two-layer structure. However, this more involved setting also leads to more aspects of reliability that must be measured. It is not the same to evaluate reliability in the choice of the semantic domain to be applied than in the chosen codes within a particular domain or the agreement achieved when distinguishing between relevant and irrelevant matter. To assess the reliability of these aspects, several variants of Krippendorff's $\alpha$ have being proposed in the literature (up to 10 are mentioned in \cite[Section 12.2.3]{Krippendorff:2018}). In this vein, we explain some of these variants and how they can be reduced to the universal $\alpha$ coefficient of Section \ref{sec:universal-alpha} after an algorithmic translation by re-labeling codes. As Section \ref{sec:universal-alpha}, this framework is an extension of the authors' work \cite{devops}.

To be precise, in coding we usually need to consider a two-layers setting as follows. First, we have a collection of $s > 1$ semantic domains, $S_1, \ldots, S_s$. A semantic domain defines a space of distinct concepts that share a common meanings (say, $S_i$ might be colors, brands, feelings...). Subsequently, each semantic domain embraces mutually exclusive concepts indicated by a code. Hence, for $1 \leq i \leq s$, the domain $S_i$ decomposes into $r_i \geq 1$ codes, that we denote by $C_1^i, \ldots, C_{r_i}^i$. For design consistency, these semantic domains must be logically and conceptually independent. This principle translates into the fact that there exists no shared codes between different semantic domains and two codes within the same semantic domain cannot be applied at the same time by a coder.

Now, the data under analysis (e.g. scientific literature, newspapers, videos,  interviews) is chopped into items, which in this context are known as \textit{quotations}, that represent meaningful parts of the data by their own. The decomposition may be decided by each of the coders (so different coders may have different quotations) or it may be pre-established (for instance, by the codebook creator or the designer of the ICA study). In the later case, all the coders share the same quotations so they cannot modify their limits and they should evaluate each quotation as a block. In order to enlighten the notation, we will suppose that we are dealing with this case of pre-established quotations. Indeed, from a mathematical point of view, the former case can be reduced to this version by refining the data subdivision of each coder to get a common decomposition into the same pieces.

Therefore, we will suppose that the data is previously decomposed into $m \geq 1$ items or quotations, $I_1, \ldots, I_m$. Observe that the union of all the quotations must be the whole matter so, in particular, irrelevant matter is also included as quotations. Now, each of the coders $J_\alpha$, $1 \leq \alpha \leq n$, evaluates the quotations $I_\beta$, $1 \leq i \leq m$, assigning to $I_\beta$ any number of semantic domains and, for each chosen semantic domain, one and only one code. No semantic domain may be assigned in the case that the coder considers that $I_\beta$ is irrelevant matter, and several domains can be applied to $I_\beta$ by the same coder.

Hence, as byproduct of the evaluation process, we obtain a collection of sets $\Sigma = \left\{\sigma_{\alpha, \beta}\right\}$, for $1 \leq \alpha \leq n$ and $1 \leq \beta \leq m$. Here, $\sigma_{\alpha, \beta} = \left\{C_{j_1}^{i_1}, \ldots, C_{j_p}^{i_p}\right\}$ is the collection of codes that the coder $J_\alpha$ assigned to the quotation $I_\beta$. The exclusion principle of the codes within the semantic domain means that the collection of chosen semantic domains $i_1, \ldots, i_p$ contains no repetitions.

\begin{rmk}
\label{rmk:computation-ica-length}
To be precise, as proposed in \cite{Krippendorff:1995}, when dealing with a continuum of matter each of the quotations must be weighted by its length in the observed and expected coincidences matrices. This length is defined as the amount of atomic units the quotation has (say characters in a text or seconds in a video). In this way, (dis)agreements in long quotations are more significant than (dis)agreements in short quotations. This can be easily incorporated to our setting just by refining the data decomposition to the level of units. In this way, we create new quotations having the length of an atomic unit. Each new atomic quotation is judged with the same evaluations as the old bigger quotation. In the coefficients introduced below, this idea has the mathematical effect that, in the sums of Equation (\ref{eq:disagreement}), each old quotation appears as many times as atomic units it contains, which is the length of such quotation. Therefore, in this manner, the version explained here computes the same coefficient as in \cite{Krippendorff:1995}.
\end{rmk}

In order to quantify the degree of agreement achieved by the coders in the evaluations $\Sigma$, several variants of Krippendorff's $\alpha$ are proposed in the literature \cite{Krippendorff:2016,Krippendorff:2018}. Some of the most useful for case studies, and the ones implemented in Atlas.ti, are the following variants.

\begin{itemize}
\item The coefficient $\globalalphabin$: This is a global measure. It quantifies the agreement of the coders when identifying relevant matter (quotations that deserve to be coded) and irrelevant matter (part of the corpus that is not coded).

\item The coefficient $\alphabin$: This coefficient is computed on a specific semantic domain $S_i$. It is a measure of the degree of agreement that the coders achieve when choosing to apply a semantic domain $S_i$ or not.

\item The coefficient $\cualpha$: This coefficient is computed on a semantic domain $S_i$. It indicates the degree of agreement to which coders identify codes within $S_i$.

\item The coefficient $\Cualpha$: This is a global measure of the goodness of the partition into semantic domains. $\Cualpha$ measures the degree of reliability in the decision of applying the different semantic domains, independently of the chosen code.
\end{itemize}

\begin{figure}[ht]
\centering
\includegraphics [width=14cm]{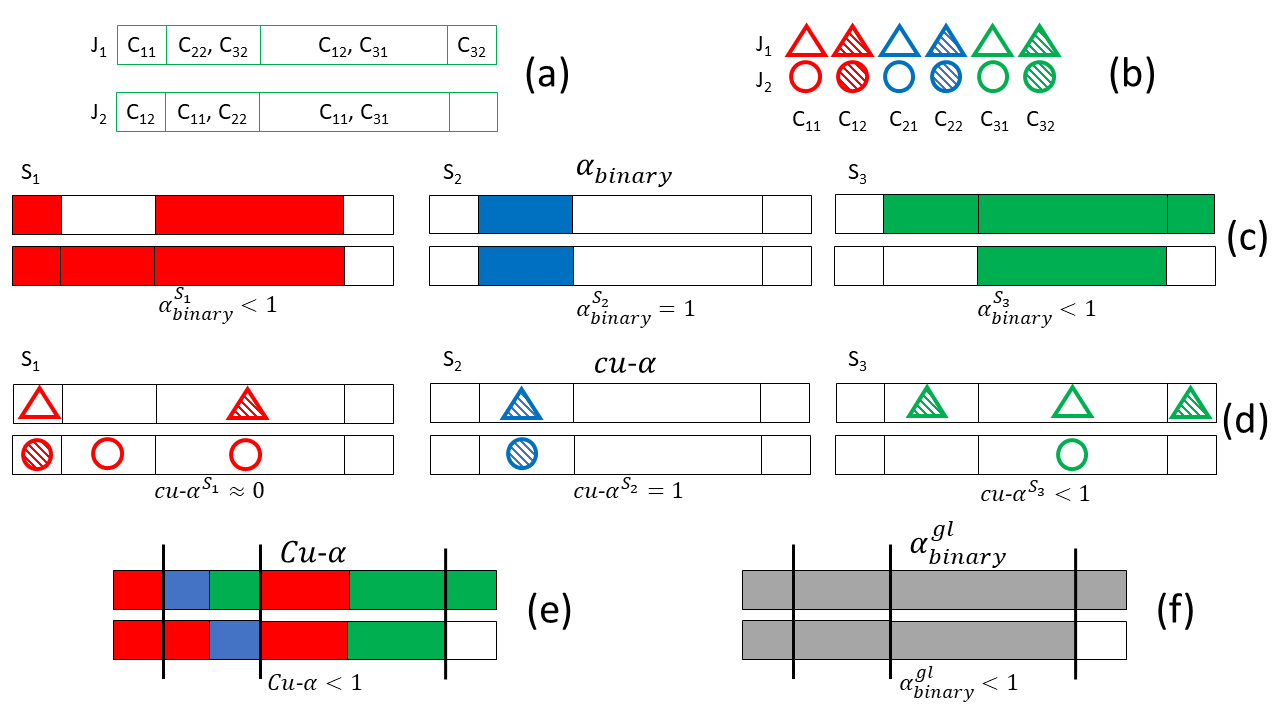}
\caption{Illustrative example for coefficients}
\label{fig:7}
\end{figure}

Before diving into the detailed formulation, let us work out an illustrative example. Figure~\ref{fig:7} shows an example of the use of these coefficients. Let us consider three semantic domains, which their respective codes being as follows
$$
	S_1 = \left\{C_{11}, C_{12}\right\}, \quad S_2 = \left\{C_{21}, C_{22}\right\}, \quad S_3 = \left\{C_{31}, C_{32}\right\}.
$$

The two coders, $J_1$ and $J_2$, assign codes to four quotations as shown in Figure~\ref{fig:7}(a). We created a graphical metaphor so that each coder, each semantic domain, and each code are represented as shown in Figure~\ref{fig:7}(b). Each coder is represented by a shape, so that $J_1$ is represented by triangles and $J_2$ by circles. Each domain is represented by a colour, so that $S_1$ is red, $S_2$ is blue and $S_3$ is green. Each code within the same semantic domain is represented a fill, so that ${C_{i1}}$ codes are represented by a solid fill and ${C_{i2}}$ codes are represented by dashed fill.

The coefficient $\alphabin$ is calculated per domain (i.e. $S_1$ red, $S_2$ blue, $S_3$ green) and analyzes whether the coders assigned or not a domain---independently of the code---to the quotations (see Figure~\ref{fig:7}(c)). Notice that we only focus on the presence or absence of a semantic domain by quotation, so Figure~\ref{fig:7}(c) only takes into account the color. Now, the $\alphabin$ coefficient measures the agreement that the coders achieved in assigning the same color to the same quotation. The bigger the coefficient, the better the agreement. In this way, we get total agreement ($\alphabin=1$) for $S_2$ as both coders assigned this domain (blue) to the second quotation and the absence of this domain in the rest of quotations. On the other hand, $\alphabin<1$ for $S_1$ as $J_1$ assigned this domain (red) to quotations 1 and 3 while $J_2$ assigned it to quotations 1, 2 and 3, leading to a disagreement in quotation 2.

The coefficient $\cualpha$ is also calculated per domain (i.e. $S_1$ red, $S_2$ blue, $S_3$ green), but it measures the agreement attained when applying the codes of that domain. In other words, given a domain $S_i$, this coefficient analyzes whether the coders assigned the same codes of $S_i$ (i.e.\ the same fills) to the quotations or not. In this way, as shown in Figure~\ref{fig:7}(d), it only focuses on the applied fills to each quotation. In particular, observe that $\cualpha=1$ for $S_2$ since both coders assigned the same code to the second quotation and no code from this domain to the rest of quotations, i.e. total agreement. Also notice that $\cualpha<1$ for $S_3$ as the coders assigned the same code of $S_3$ to the third quotation but they did not assign the same codes of $S_3$ to the rest of quotations. Finally, observe that cu-alpha for $S_1$ is very small (near to zero) since the coders achieve no agreement on the chosen codes.

With respect to the global coefficients, the coefficient $\Cualpha$ analyzes all the domains as a whole, but it does not take into account the codes within each domain. In this way, in Figure~\ref{fig:7}(e), we colour each segments with the colors corresponding to the applied semantic domain (regardless of the particular used code). From these chromatic representation, $\Cualpha$ measures the agreement in applying these colours globally between the coders. In particular, notice that $\Cualpha<1$ as both coders assigned the same domain $S_1$ to the first quotations and the domains $S_1$ and $S_3$ to the third quotation, but they did not assign the same domains in the second and fourth quotations.

Finally, the $\globalalphabin$ coefficient measures the agreement in the selection of relevant matter, as shown in Figure~\ref{fig:7}(f). In this case, both coders recognized the first three segments as relevant (they were coded), as highlighted in gray in the figure. However, $J_2$ considered that the forth quotation was irrelevant (it was not coded), as marked in white, and $J_1$ marked it as relevant, in gray. In this way, we have that $\globalalphabin < 1$.

\subsection{The coefficient $\globalalphabin$}
\label{sec:globalalphabin}

The first variation of Krippendorff's $\alpha$ coefficient that we consider is the $\globalalphabin$ coefficient. It is a global measure that summarizes the agreement of the coders for recognizing relevant parts of the matter. For computing it, we consider a set of labels have only two labels, that semantically represent `recognized as relevant' (1) and `not recognized as relevant' (0). Hence, we take
$$
    \Lambda = \left\{1, 0\right\}.
$$

Now, using the whole set of evaluation $\Sigma$, we create a new labelling $\Omega_{bin}^{gl} = \left\{\omega_{\alpha, \beta}\right\}$ as follows. Let $1 \leq \alpha \leq n$ and $1 \leq \beta \leq m$. We set $\omega_{\alpha,\beta} = \left\{1\right\}$ if the coder $J_\alpha$ assigned some code to the quotation $I_\beta$ (i.e.\ if $\sigma_{\alpha, \beta} \neq \emptyset$) and $\omega_{\alpha,\beta} = \left\{0\right\}$ otherwise (i.e.\ if $J_\alpha$ did not code $I_\beta$, that is $\sigma_{\alpha, \beta} = \emptyset$). From this set of evaluations, $\Omega_{bin}^{gl} = \left\{\omega_{\alpha, \beta}\right\}$, $\globalalphabin$ is given as
$$
    \globalalphabin = \alpha(\Omega_{bin}^{gl}).
$$

Therefore, $\globalalphabin$ measure the degree of agreement that the coders achieved when recognizing relevant parts, that is coded parts, and irrelevant matter. A high value of $\globalalphabin$ may be interpreted as that the matter is well structured and it is relatively easy to detect and isolate the relevant parts of information.

\begin{rmk}
In many studies (for instance in case studies in software engineering), it is customary that a researcher pre-processes the raw data to be analyzed, say by transcribing it or by writing it down into a ICA software like Atlas.ti. In that case, usually this pre-processor selects the parts that must be analyzed and chops the matter into quotations before the starting of the judgement process. In this way, the coders are required to code these pre-selected parts, so that they no longer chop the matter by themselves and they code all the quotations. Hence, we always get that $\globalalphabin = 1$, since the evaluation protocol forces the coders to consider as relevant matter the selected parts by the pre-processor. Therefore, in these scenarios, the $\globalalphabin$ coefficient is not useful for providing reliability on the evaluations and other coefficients of the $\alpha$ family are required.
\end{rmk}

\subsection{The coefficient $\alphabin$}
\label{sec:alphabin}

The second variation of the Krippendorff's $\alpha$ coefficient is the so-called $\alphabin$ coefficient. This is a coefficient that must be computed on a specific semantic domain. Hence, let us fix a semantic domain $S_{i}$ for some fixed $i$ with $1 \leq i \leq s$. As above, the set of labels will have only two labels, that semantically represent `voted $S_{i}$' (1) and `did not vote $S_{i}$' (0). Hence, we take
$$
    \Lambda = \left\{1, 0\right\}.
$$

For the assignment of labels to items, the rule is as follows. For $1 \leq \alpha \leq n$ and $1 \leq \beta \leq m$, we set $\omega_{\alpha,\beta} = \left\{1\right\}$ if the coder $J_\alpha$ assigned some code of $S_{i}$ to the quotation $I_\beta$ (i.e.\ if $C_j^{i} \in \sigma_{\alpha, \beta}$ for some $1 \leq j \leq r_{i}$) and $\omega_{\alpha,\beta} = \left\{0\right\}$ otherwise. Observe that, in particular, $\omega_{\alpha,\beta} = \left\{0\right\}$ if $J_\alpha$ considered that $I_\beta$ was irrelevant matter. From this set of evaluations, $\Omega_{binary}^{S_{i}} = \left\{\omega_{\alpha, \beta}\right\}$, $\alphabin^{S_i}$ is given as
$$
    \alphabin^{S_{i}} = \alpha(\Omega_{binary}^{S_{i}}).
$$

In this way, the coefficient $\alphabin^{S_{i}}$ can be seen as a measure of the degree of agreement that the coders achieved when choosing to apply the semantic domain $S_{i}$ or not. A high value of $\alphabin^{S_{i}}$ is interpreted as an evidence that the domain $S_{i}$ is clearly stated, its boundaries are well-defined and, thus, the decision of applying it or not is near to be deterministic. However, observe that it does not measure the degree of agreement in the application of the different codes within the domain $S_{i}$. Hence, it may occur that the boundaries of the domain $S_{i}$ are clearly defined but the inner codes are not well chosen. This is not a task of the $\alphabin^{S_{i}}$ coefficient, but of the $\cualpha^{S_{i}}$ coefficient explained below.

\begin{rmk}
By the definition of $\alphabin$, in line with the implementation in Atlas.ti \cite{Atlas:2019}, the irrelevant matter plays a role in the computation. As we mentioned above, all the matter that was evaluated as irrelevant (i.e.\ was not coded) is labelled with $\left\{0\right\}$. In particular, a large corpus with only a few sparse short quotations may distort the value of $\alphabin$.
\end{rmk}

\subsection{The coefficient $\cualpha$}
\label{sec:cu-alpha}

Another variation of the Krippendorff's $\alpha$ coefficient is the so-called $\cualpha$ coefficient. As the previous variation, this coefficient is computed per semantic domain, say $S_{i}$ for some $1 \leq i \leq s$. Suppose that this semantic domain contains codes $C^{i}_1, \ldots, C^{i}_r$. The collection of labels is now a set
$$
    \Lambda = \left\{\cC_1, \ldots, \cC_r\right\}.
$$
Semantically, they are labels that represent the codes of the chosen domain $S_{i}$.

For the assignment of labels to items, the rule is as follows. For $1 \leq \alpha \leq n$ and $1 \leq \beta \leq m$, we set $\omega_{\alpha,\beta} = \cC_k$ if the coder $J_\alpha$ assigned the code $C_k^{i}$ of $S_{i}$ to the item (quotation) $I_\beta$. Recall that, from the exclusion principle for codes within a semantic domain, the coder $J_\alpha$ applied at most one code from $S_{i}$ to $I_\beta$. If the coder $J_\alpha$ did not apply any code of $S_{i}$ to $I_\beta$, we set $\omega_{\alpha, \beta} = \emptyset$. From this set of judgements $\Omega_{cu}^{S_{i}} = \left\{\omega_{\alpha, \beta}\right\}$, $\cualpha^{S_i}$ is given as
$$
    \cualpha^{S_{i}} = \alpha(\Omega_{cu}^{S_{i}}).
$$

\begin{rmk}
\label{rmk:computation-alpha-non-voted}
As explained in Remark \ref{ref:not-voted-item}, for the computation of the observed and expected coincidence matrices, only items that received at least two evaluations with codes of $S_{i}$ from two different coders count. In particular, if a quotation is not evaluated by any coder (irrelevant matter), received evaluations for other domains but not for $S_{i}$ (matter that does not corresponds to the chosen domain) or only one coder assigned to it a code from $S_{i}$ (singled-voted), the quotation plays no role in $\cualpha$. This limitation might seem a bit cumbersome, but it could be explained by arguing that the presence/absence of $S_{i}$ is measured by $\alphabin^{S_{i}}$ so it will be redundant to take it into account for $\cualpha^{S_{i}}$ too.
\end{rmk}

\subsection{The coefficient $\Cualpha$}
\label{sec:Cu-alpha}

The last variation of Krippendorff's $\alpha$ coefficient that we consider in this study is the so-called $\Cualpha$ coefficient. In contrast with the previous coefficients, this is a global measure of the goodness of the partition into semantic domains. Suppose that our codebook determines semantic domains $S_1, \ldots, S_s$. In this case, the collection of labels is the set
$$
    \Lambda = \left\{\cS_1, \ldots, \cS_s\right\}.
$$
Semantically, they are labels representing the semantic domains of our codebook.

We assign labels to items as follows. Let $1 \leq \alpha \leq n$ and $1 \leq \beta \leq m$. Then, if $\sigma_{\alpha, \beta} = \left\{C_{j_1}^{i_1}, \ldots, C_{j_p}^{i_p}\right\}$, we set $\omega_{\alpha, \beta} = \left\{\cS_{i_1}, \ldots, \cS_{i_p}\right\}$. In other words, we label $I_\beta$ with the labels corresponding to the semantic domains chosen by coder $J_\alpha$ for this item, independently of the particular code. Observe that this is the first case in which the final evaluation $\Omega$ might be multivaluated. From this set of judgements, $\Omega_{Cu} = \left\{\omega_{\alpha, \beta}\right\}$, $\Cualpha$ is given as
$$
    \Cualpha = \alpha(\Omega_{Cu}).
$$

In this way, $\Cualpha$ measures the degree of reliability in the decision of applying the different semantic domains, independently of the particular chosen code. Therefore, it is a global measure that quantifies the logical independence of the semantic domains and the ability of the coders of looking at the big picture of the matter, only from the point of view of semantic domains.


\section{Atlas.ti for Inter-Coder Agreement (ICA): a tutorial}
\label{sec:atlas}

In this section, we describe how to use the ICA utilities provided by Atlas.ti v8.4\footnote{During the final phase of this work, Atlas.ti v9 has been released. The screenshots and commands mentioned in this paper correspond to v8.4. However, regarding the calculation of the ICA coefficients, the new version does not provide new features. In this way, the tutorial shown in this section is also valid for v9, with only minimum changes in the interface.} \cite{Atlas:2019} (from now on, shortened as Atlas) as well as a guide for interpreting of the obtained results. We will assume that the reader is familiar with the general operation of Atlas and focus on the computation and evaluation of the different ICA coefficients calculated by Atlas. 

This section is structured as follows. First, in Section \ref{sec:atlas-coefficients} we describe the different operation methods provided by Atlas for the computation of ICA. In Section \ref{sec:atlas-case-study}, we describe briefly the protocol for analyzing case study research involving qualitative data analysis in software engineering and we introduce a running example on the topic that will serve as a guide along all the tutorial. Finally, in Section \ref{sec:atlas-computation-ica}, we discuss the calculation, interpretation and validity conclusions that can be drawn from the ICA coefficients provided by Atlas.

\subsection{Coefficients in Atlas for the computation of ICA}
\label{sec:atlas-coefficients}

Atlas provides three different methods for computing the agreement between coders, namely simple percent agreement, Holsti Index (both can be checked in Section \ref{sec:background-percent-agreement}), and Krippendorff's $\alpha$ coefficients (see Sections \ref{sec:background-krip-alpha} and \ref{sec:theoretical-framework}).


\subsubsection{Simple percent agreement and Holsti Index}
\label{sec:atlas-percent}

As we pointed out in Section \ref{sec:background-percent-agreement}, it is well reported in the literature that simple percent agreement is not a valid ICA measure, since it does not take into account the agreement that the coders can be attained by chance.

On the other hand, the Holsti Index \cite{holsti1969content}, as referred in Section \ref{sec:background-percent-agreement}, is a variation of the percent agreement that can be applied when there are no pre-defined quotations and each coder selects the matter that considers relevant. Nevertheless, as in the case of the percent agreement, it ignores the agreement by chance so it is not suitable for a rigorous analysis.

In any case, Atlas provides us these measures that allow us to glance at the results and to get an idea of the distribution of the codes. However, they should not be used for drawing any conclusions about the validity of the coding. For this reason, in this tutorial we will focus on the application and interpretation of Krippendorff's $\alpha$ coefficients.

\subsubsection{Krippendorff's $\alpha$ in Atlas}
\label{sec:atlas-Krippendorff}

Atlas also provides an integrated method for computing the Krippendorff's $\alpha$ coefficients. However, it may be difficult at a first sight to identify the prompted results since the notation is not fully consistent between Atlas and some reports in the literature. The development of the different versions of the $\alpha$ coefficient has taken around fifty years and, during this time, the notation for its several variants has changed. Now, a variety of notations co-exists in the literature that may confuse the unfamiliar reader with this ICA measure.

In order to clarify these relations, in this paper we always use the notation introduced in Section \ref{sec:theoretical-framework}. These notations are based on the ones provided by Atlas, but some slightly differences can be appreciated. For the convenience of the reader, in Table \ref{tab:equivalence-notations} we include a comparative between the original Krippendorff's notation, the Atlas notation for the $\alpha$ coefficient and the notation used in this paper.

\begin{rmk}
Empirically, we have discovered that the semantics that the software Atlas applies for computing the coefficients $\alphabin$/$\globalalphabin$ and $\cualpha$/$\Cualpha$ are the ones explained in this paper, as provided in Section \ref{sec:theoretical-framework}.
\end{rmk}

\begin{table}[h]
    \centering
\small
    \begin{tabular}{c|c|c|c}
        Name & Krippendorff's \cite{Krippendorff:2018} & Atlas \cite{Atlas:2019,friese:2019} & This paper \\\hline
        Global binary $\alpha$ & $_{u}\alpha$ & alpha-binary (global) & $\globalalphabin$  \\
        Binary $\alpha$ per domain & $_{|cu}\alpha$, $_{|u}\alpha$ & alpha-binary (domain) & $\alphabin$ \\
        cu-$\alpha$ & $_{(s)u}\alpha$, $_{(k)u}\alpha$ & cu-alpha & $\cualpha$ \\
        Cu-$\alpha$ & $_{Su}\alpha$ & Cu-alpha & $\Cualpha$ \\
    \end{tabular}
    \caption{Equivalence of notations between the variants of the $\alpha$ coefficient}
    \label{tab:equivalence-notations}
\end{table}
\vspace{-0.3cm}
\normalsize

\subsection{Case study: instilling DevOps culture in software companies}
\label{sec:atlas-case-study}

This tutorial uses as guiding example an excerpt of a research conducted by the authors in the domain of DevOps \cite{devops}. The considered example is an exploratory study to characterize the reasons why companies move to DevOps and what results do they expect to obtain when adopting the DevOps culture \cite{leite2019survey}. This exploratory case study is based on interviews to software practitioners from 30 multinational software-intensive companies. The study has been conducted according to the guidelines for performing qualitative research in software engineering proposed by Wohlin et al.\ \cite{wohlin2012experimentation}.

\begin{figure}[!h]
\centering
\includegraphics [width=0.51\textwidth]{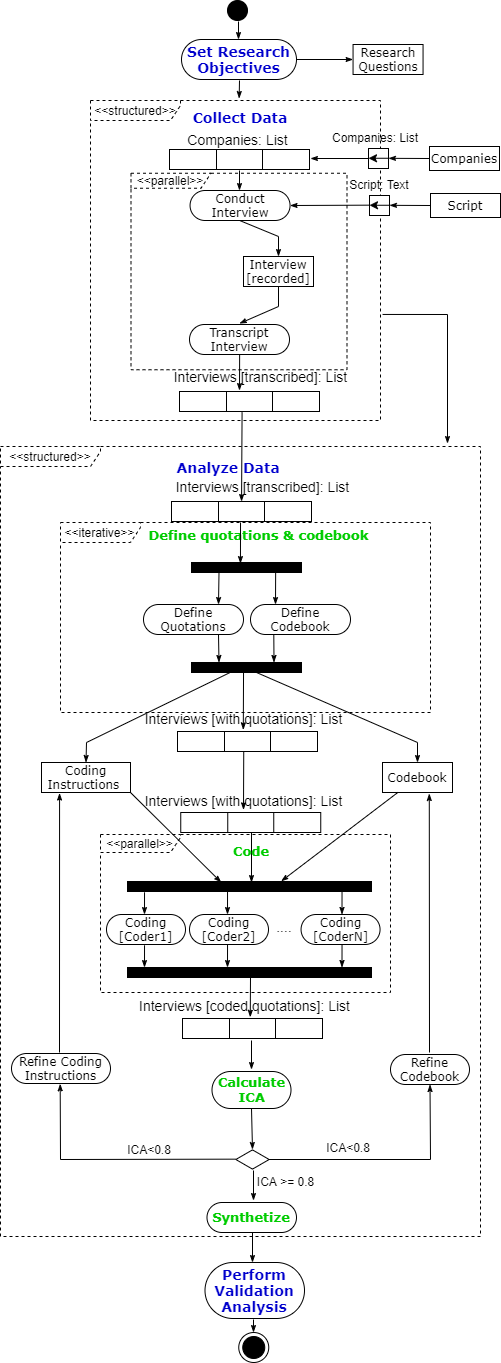}
\caption{Phases for conducting case study research involving qualitative data analysis in software engineering}
\label{figure:2}
\end{figure}

In Figure \ref{figure:2} we show, through a UML activity diagram \cite{rumbaugh1999unified}, the different stages that comprise the above-mentioned study. For the sake of completeness, in the following exposition each step of the analysis is accompanied with a brief explanation of the underlying qualitative research methodology that the authors carried out. For a complete description of the methodology in qualitative research and thematic analysis, please check the aforementioned references.

\subsubsection{Set research objectives}

The first step needed for conducting an exploratory study is to define the aim of the prospective work, the so-called research question (RQ). These objectives must be clearly stated and the boundaries of each research question should be undoubtedly demarcated.

In the case study of the running example presented in this paper, we propose one research question related to the implications of instilling a DevOps culture in a company, which is the main concern of the analysis.
\begin{center}
    \textit{ \textbf{RQ:} What problems do companies try to solve by implementing DevOps?}
\end{center}

\subsubsection{Collect data}

The next step in the research is to collect the empirical evidences needed for understanding the phenomenon under study. Data is the only window that the researchers have to the object of research, so getting high quality data typically leads to good researches.
As a rule-of-thumb, the better the data, the more precise the conclusions can be drawn. 

There are two main methods for collecting information in qualitative analysis, and both are particularly useful in software engineering: questionnaires and interviews \cite{wohlin2012experimentation}. Usually, questionnaires are easier to issue, since they can be provided by email or web pages that can be access whenever is preferable for the person in charge of answering it. On the other hand, interviews tent to gather a more complete picture of the phenomenon under study since there exists an active interaction between interviewer and interviewee, typically face to face. In this way, interviews are usually better suited for case studies since they allow the researcher to modify the questions to be asked on the fly, in order to emphasize the key points under analysis. As a drawback, typically the number of answers that can be obtained through a questionnaire is much larger than the number of interviews that can be conducted, but the later usually lead to higher quality data.

In the study considered in this paper, the data collection method was semi-structured interviews to software practitioners of 30 companies. The interviews were conducted face-to-face, using the Spanish language, and the audio was recorded with the permission of the participants, transcribed for the purpose of data analysis, and reviewed by respondents. In the transcripts, the companies were anonymized by assigning them an individual identification number from ID01 to ID30. The full script of the interview is available at the project's webpage\footnote{ \href{https://blogs.upm.es/devopsinpractice}{https://blogs.upm.es/devopsinpractice}.}.

\subsubsection{Analyze data}
\label{sec:case-study-data-analysis}

This is the most important phase in the study. In this step, the researchers turn the raw data into structured and logically interconnected conclusions. On the other hand, due to its creative component, it is the less straightforward phase in the cycle.

To help researchers to analyze the data and to draw the conclusions, there exists several methods for qualitative data analysis that can be followed. In the DevOps exploratory study considered here, the authors conducted a thematic analysis approach \cite{cruzes2011recommended,thomas2008methods}. Thematic analysis is a method for identifying, analyzing, and reporting patterns within the data. For that purpose, the data is chopped into small pieces of information, the quotations or segments, that are minimal units of data. Then, some individuals (typically some of the researchers) act as coders, codifying the segments to highlight the relevant information and to assign it a condensate description, the code. In the literature, codes are defined as ``descriptive labels that are applied to segments of text from each study'' \cite{cruzes2011recommended}. In order to easy the task of the coders, the codes can be grouped into bigger categories that share come higher level characteristics, forming the semantic domains (also known as themes in this context). This introduce a multi-level coding that usually leads to richer analysis.

A very important point is that splitting of the matter under study into quotations can be provided by a non-coder individual (typically, the thematic analysis designer), or it can be a task delegated to the coders. In the former case, all the coders work with the same segments, so it is easier to achieve a high level of consensus that leads to high reliability in the results of the analysis. In the later case, the coders can decide by themselves how to cut the stream of data, so hidden phenomena can be uncovered. However, the cuts may vary from a coder to another, so there exists a high risk of getting too diverse codings that cannot be analyzed under a common framework. 

Thematic analysis can be instrumented through Atlas \cite{Atlas:2019,friese:2019}, which provides an integrated framework for defining the quotations, codes and semantic domains, as well as for gathering the codings and to compute the attained ICA.

In the study considered in this section, the method for data analysis followed is described in the four phases described below (see also Figure \ref{figure:2}).

\begin{enumerate}
    \item \textbf{Define quotations \& codebook.} In the study under consideration, the coders used pre-defined quotations. In this way, once the interviews were transcripted, researcher R1 chopped the data into its unit segments that remain unalterable during the subsequent phases. In parallel, R1 elaborated a codebook by collecting all the available codes and their aggregation into semantic domains. After completing the codebook, R1 also created a guide with detailed instructions about how to use the codebook and how to apply the codes.
    
    The design of the codebook is accomplished through two different approaches: a deductive approach \cite{miles1994qualitative} for creating semantic domains and an inductive approach  \cite{corbin2008techniques} for creating codes. In the first phase, the deductive approach, R1 created a list of of semantic domains in which codes will be grouped inductively during the second phase. These initial domains integrate concepts known in the literature. Each domain is named P01, P02, P03, etc. Domains were written with uppercase letters (see Figure \ref{figure:3}).
    
    \begin{figure}[!h]
\centering
\includegraphics [width=0.5\textwidth]{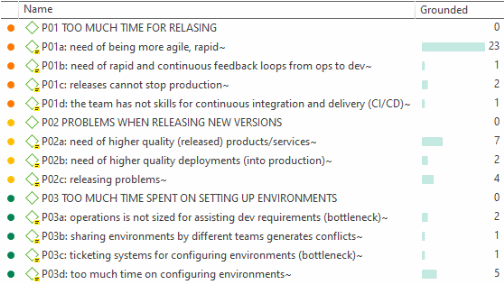}
\caption{Atlas code manager}
\label{figure:3}
\end{figure}
    
    In the second phase, the inductive approach, R1 approached the data (i.e.\ the interviews' transcriptions) with the research question RQ in mind. R1 reviewed the data line by line and created the quotations. R1 also assigned them a code (new or previously defined) in order to get a comprehensive list of all the needed codes. As more interviews were analyzed, the resulting codebook was refined by using a constant comparison method that forced R1 to go back and forth.
    
    Additionally, the codes were complemented with a brief explanation of the concept they describe. This allows R1 to guarantee that the collection of created codes satisfy the requirements imposed by thematic analysis, namely exhaustiveness and mutual exclusiveness. The exhaustiveness requirement means that the codes of the codebook must cover all the relevant aspects for the research. Mutual exclusiveness means that there must exist no overlapping in the semantics of each code within a semantic domain. In this way, the codes of a particular semantic domain must capture disjoint aspects and complementary aspects, which implies that the codes should have explicit boundaries so that they are not interchangeable or redundant. This mutual exclusiveness translates into the fact that, during the coding phase, a coder cannot apply several codes of the same semantic domain to the same quotation. In other words, each coder can apply at most a code of each semantic domain to each quotation.
    

\item \textbf{Code.} In this phase, the chosen coders (usually researchers different than the codebook designer) analyze the prescribed quotations created during phase (1). For that purpose, they use the codebook as a statement of the available semantic domains and codes as well as the definitions of each one, scope of application and boundaries. It is crucial for the process that the coders apply the codes exactly as described in the codebook. No modifications on the fly or alternative interpretations are acceptable.

Nevertheless, the coders are encourage to annotate any problem, diffuse limit or misdefinition they find during the coding process. After the coding process ends, if the coders consider that the codebook was not clear enough or the ICA measured in phase (3) does not reach an acceptable level, the coders and the codebook designer can meet to discuss the found problems. With this information, the codebook designer creates a new codebook and instructions for coding that can be used for a second round of codings. This iterative process can be conducted as many times as needed until the coders consider that the codebook is precise enough and the ICA measures certify an acceptable amount of reliability.

In the case study of ICA considered in this paper, the coding process involved two researchers different than R1, that acted as coders C1 and C2. They coded the matter according to the codebook created by R1.

\item \textbf{Calculate ICA.} It is a quite common misconception in qualitative research that no numerical calculations can be performed for the study. Qualitative research aims to understand very complex an unstructured phenomena, for which a semantic analysis of the different facets and their variations is required. However, by no means this implies that no mathematical measures can be obtained for controlling the process. Due to its broad and flexible nature, qualitative research is highly sensible to introduce biases in the judgements of the researchers, so it is mandatory to supervise the research through some reliability measure that are usually numerical \cite{Krippendorff:2018}. In this way, the quantitative approach takes place in a higher level, as meta-analysis of the conducted process in order to guarantee mathematical reliability in the drawn conclusions. Only when this formal quality assurance process is satisfactory, researchers can trust in the conclusions and the method is sound and complete.

Therefore, to avoid biases and be confident that the codes mean the same to anyone who uses them, it is necessary to build that confidence. According to Krippendorff \cite{Krippendorff:2018}, reliability grounds this confidence empirically and offers the certainty that research findings can be reproduced.

In the presented example about a DevOps case study, we used Inter-Coder Agreement (ICA) analysis techniques for testing the reliability of the obtained codebook. In this way, after coding, another researcher, R4, calculated and interpreted the ICA between C1 and C2. If coders did not reach an acceptable level of reliability, R1 analyzes the disagreements pointed out by R4 to find out why C1 and C2 had not understood a code in the same mode. 

Using this acquired knowledge, R1 delivers a refined new version of the codebook and the accompanying use instructions. R1 also reviews the coding of those quotations that led to disagreement between C1 and C2, modifying it according to the new codebook when necessary. Notice that, if a code disappears in the new version of the codebook, it also must disappear of all the quotations that were asigned with it.

At this point, C1 and C2 can continue coding on a new subset of interviews' transcriptions. This process is repeated until the ICA reached an acceptable level of reliability (typicall $\geq 0.8$). In Section \ref{sec:atlas-computation-ica} it is provided a detailed explanation about how to compute and interpret ICA coeffients in Atlas.

\item \textbf{Synthetize.} Once the loop (1)-(2)-(3) has been completed because the ICA measures reached an acceptable threshold, we can rely in the output of the coding process and start drawing conclusions. At this point, there exists a consensus about the meaning, applicability and limits of the codes and semantic domains of the codebook. 

Using this processed information, this phase aims to provide a description of higher-order themes, a taxonomy, a model, or a theory. The first action is to determine how many times each domain appears in the data in order to estimate its relevance (grounded) and to support the analysis with evidences through quotations from the interviews. After that, the co-occurrence table between semantic units should be computed, that is, the table that collects the number of times a semantic domain appears jointly with the other domains. With these data, semantic networks can be created in order to portray the relationships between domains (association, causality, etc.) as well as the relationship strength based on co-occurrence. These relations determine the density of the domains, i.e.\ the number of domains you have related to each domain. If further information is needed, it is possible to repeat these actions for each code within a domain. 


All these synthesis actions are not the main focus of this paper, so we will not describe them further. For more information and techniques, please refer to \cite{wohlin2012experimentation}.

\end{enumerate}

\subsubsection{Perform validation analysis}

As a final step, it is necessary to discuss in which way the obtained analysis and drawn conclusions are valid, as well as the threats to the validity that may jeopardize the study. In the words of Wohlin \cite{wohlin2012experimentation} ``The validity of a study denotes the trustworthiness of the results, and to what extent the results are true and not biased by the researchers' subjective point of view''.

There are several strategies for approaching to the validity analysis of the procedure. In the aforementioned case study, it was followed the methodology suggested by Creswell \& Creswell \cite{creswell2017research} to improve the validity of exploratory case studies, namely data triangulation, member checking, rich description, clarify bias, and report discrepant information. Most of these methods are out of the scope of this paper and are not described further (for more information, check \cite{creswell2017research,wohlin2012experimentation}). We mainly focus on reducing authors bias by evaluating the reliability and consistency of the codebook on which the study findings are based through ICA analysis. 

\subsection{ICA calculation}
\label{sec:atlas-computation-ica}

This section describes how to perform the ICA analysis required using Atlas to assess the validity of the exploratory study described in Section \ref{sec:atlas-case-study}. For this purpose, we use the theoretical framework developed in Section \ref{sec:theoretical-framework} regarding the different variants of Krippendorff's $\alpha$ coefficient. In this way, we will monitor the evolution of the $\alpha$ coefficients along the coding process in order to assure it reaches an acceptable threshold of reliability, as mentioned in Section \ref{sec:case-study-data-analysis}.

Nevertheless, before starting the coding/evaluation protocol, it is worthy to consider two important methodological aspects, as described below.

\begin{enumerate}
    \item The number of coders. Undoubtedly, the higher the number of involved coders, the richer the coding process. Krippendorff's $\alpha$ coefficients can be applied to an arbitrary number of coders, so there exists no intrinsic limitation to this number. On the other hand, a high number of coders may introduce too many different interpretations that may difficult to reach an agreement. 
    In this way, it is important to find a fair balance between the number of coders and the time to reach agreement. For that purpose, it may be useful to take into account the number of interviews to be analyzed, its length and the resulting total amount of quotations.
    In the case study analyzed in this section, two coders, C1 and C2 were considered for coding excerpts extracted from 30 interviews.
    
    \item The extend of the coding/evaluation loop. A first approach to the data analysis process would be to let the coders codify the whole corpus of interviews, and to get an resulting ICA measure when the coding is completed. However, if the obtained ICA is below the acceptable threshold (say $0.8$), the only solution that can be given is to refine the codebook and to re-codify the whole corpus again. This is a slow and repetitive protocol that can lead to intrinsic deviations in the subsequent codings due to cognitive biases in the coders.
    
    In this way, it is more convenient to follow an iterative approach that avoids these problems and speeds up the process. In this approach, the ICA coefficient is screened on several partially completed codings. To be precise, the designer of the case study splits the interviews into several subsets. The coders process the first subset and, after that, the ICA coefficients are computed. If this value is below the threshold of acceptance ($0.8$), there exists a disagreement between the coders when applying the codebook. At this point, the designer can use the partial coding to detect the problematic codes and to offer a refined version of the codebook and the accompanying instructions. Of course, after this revision, the previously coded matter should be updated with the new codes. With this new codebook, the coders can face the next subset of interviews, in the expectation that the newer version of the codebook will lead to decrease the disagreement. This reduces the number of complete codings needed to achieve an acceptable agreement.
    
    In the case study considered as example, the first batch of interviews comprised the first 19 interviews (ID01 to ID19). The attained ICA was unsatisfactory, so the codebook designer R1 reviewed the codebook releasing a new version. With the updated codebook, the coders codified the remaining 11 interviews (ID20 to ID30) but, now, the obtained ICA pass the acceptance threshold, which evidences a high level of reliability in the evaluations.
    
    As a final remark, it is not recommendable to replace the coders during this iterative process. Despite that Krippendorff's $\alpha$ allows to exchange coders, the new coders may not share the same vision and expertice with the codebook, requiring to roll back to previous versions (c.f.\ Remark \ref{rmk:change-judges}).
\end{enumerate}

Now, we present the calculation and interpretation of each of the four coefficients mentioned in Section \ref{sec:theoretical-framework}. In order to emphasize the methodological aspects of the process, we focus only on some illustrative instances for each of the two rounds of the coding/evaluation protocol.

The first step in order to address the ICA analysis is to create an Atlas project and to introduce the semantic domains and codes compiled in the codebook. In addition, all the interviews (and their respective quotations, previously defined by R1) should be loaded as separated documents and all the codings performed by the coders should be integrated in the Atlas project. This is a straightforward process that is detailed in the Atlas' user manual \cite{Atlas:2019}.

\begin{figure}[!h]
\centering
\includegraphics [width=0.55\textwidth]{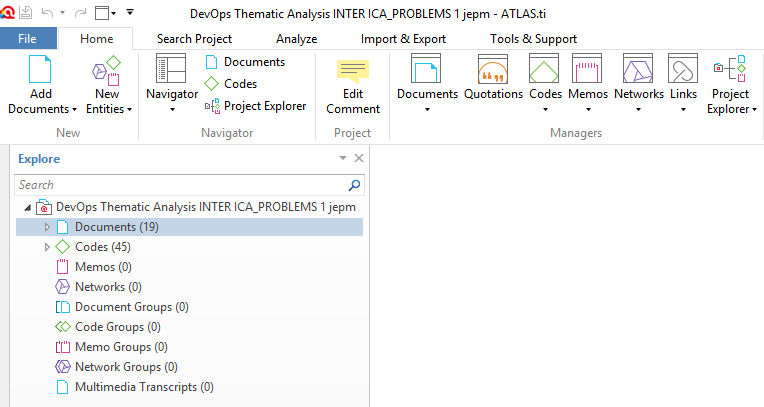}
\caption{Documents and codes for the Atlas project}
\label{figure:5}
\end{figure}

To illustrate our case study, we have a project containing the codings of the two coders for the first 19 interviews. The codebook has 10 semantic domains and 35 codes (Figure \ref{figure:5}). Observe that Atlas reports a total of 45 codes, since it treats semantic domains as codes (despite that it will work as an aggregation of codes for ICA purposes).

\begin{figure}[!h]
\centering
\includegraphics [width=13.5cm, height=7cm]{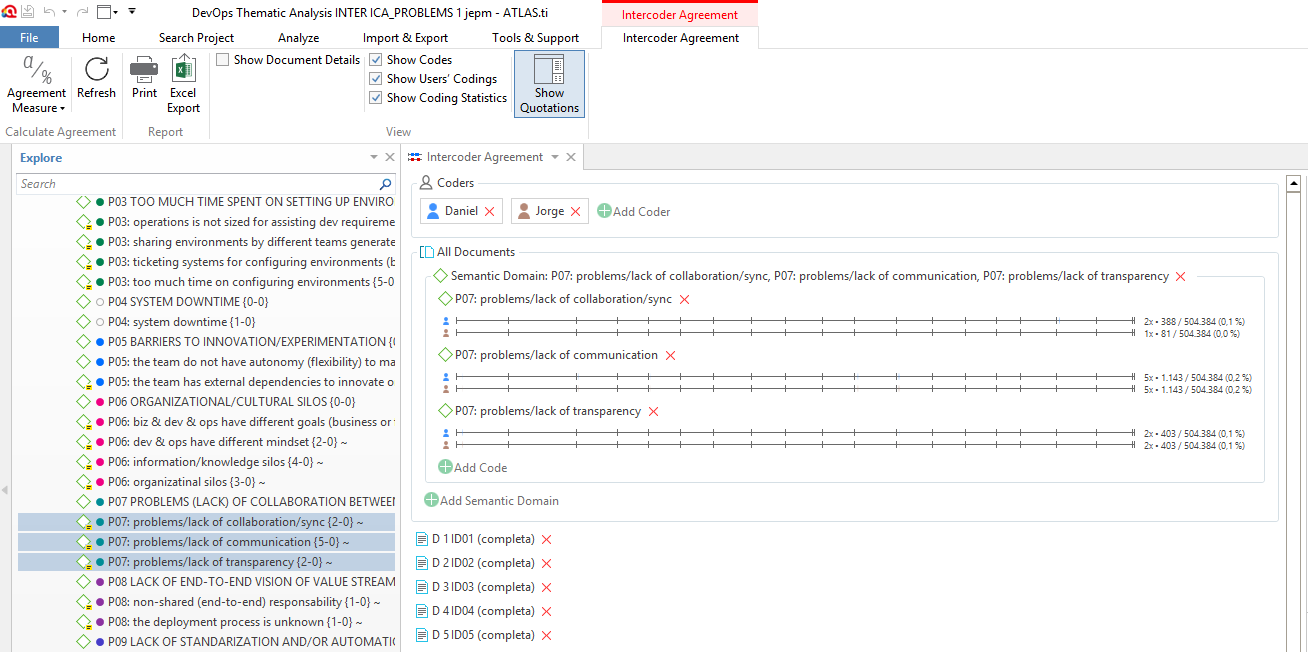}
\caption{Coders, documents and domains selected}
\label{figure:9}
\end{figure}

Now, we should select the semantic domains we wish to analyze, and the codes within them.
For example, in Figure \ref{figure:9}, the three codes associated to semantic domain P07 have been added. 

After adding a semantic domain and its codes, Atlas automatically plots a graphical representation. For each code, this graph is made of as many horizontal lines as coders (two, in our running example) that are identified with a small icon on the left. Each line is divided into segments that represent each of the documents added for analysis. 

\begin{figure}[!h]
\centering
\includegraphics [width=0.95\textwidth]{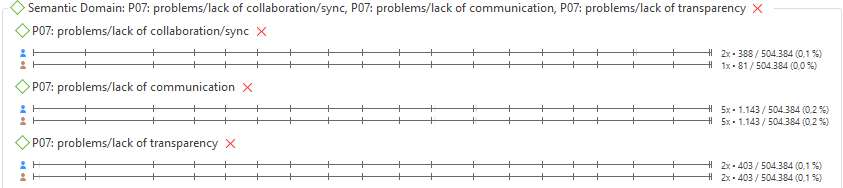}
\caption{Codification summary by code, semantic domain and coder}
\label{figure:10}
\end{figure}

As can be checked in Figure \ref{figure:10}, there are two coders (Daniel and Jorge, represented with blue and brown icons respectively) and the semantic domain P07 has three associated codes, so three groups of pairs of horizontal lines are depicted. In addition, since we selected 19 documents for this coding round, the lines are divided into 19 segments (notice that the last one is very short and it can be barely seen). Observe that the length of each segment is proportional to the total length of the file.

\begin{figure}[!h]
\centering
\includegraphics [width=0.25\textwidth]{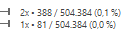}
\caption{Length information for code P07}
\label{figure:11}
\end{figure}

Moreover, on the right of the horizontal lines we find a sequence of numbers organized into two groups separated by a slash. For example, in Figure \ref{figure:11} we can see those numbers for the first code of P07 (problems/lack of collaboration/sync). The left-most group shows the number of quotations to which the corresponding code has been applied by the coder along all the documents, as well as the total length (i.e.\ the number of characters) of the chosen quotations. In the example of Figure \ref{figure:11}, the first coder (Daniel) used the first code twice, and the total length of the chosen quotations is $388$; while the second coder (Jorge) used the first code only once on a quotation of length $81$.

On the other hand, the right-most group indicates the total length of the analyzed documents (in particular, it is a constant independent of the chosen code, semantic domain or coder). This total length is accompanied with the rate of the coded quotations among the whole corpus. In this example, the total length of the documents to be analyzed is $504.384$ and the coded quotations (with a total length of $388$ and $81$ respectively) represent the $0.076\%$ and the $0.016\%$ of the corpus (rounded to $0.1\%$ and $0.0\%$ in the Atlas representation). Recall that these lengths of the coded quotations and total corpus play an important role in the computation of the $\alpha$ coefficient, as mentioned in Remark \ref{rmk:computation-ica-length}.

\begin{figure}[!h]
\centering
\includegraphics [width=0.4\textwidth]{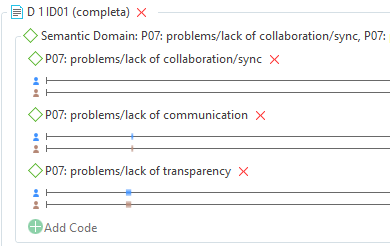}
\caption{Codified quotations in document ID01}
\label{figure:13}
\end{figure}

Each time that a coder uses a code, a small coloured mark is placed in the position of the quotation within the document. The colour of the mark agrees with the assigned color to the coder, and its length corresponds to the length of the coded quotation. Due to the short length of the chosen quotations in Figure \ref{figure:10} they are barely seen, but we can zoom in by choosing the \texttt{Show Documents Details} in the Atlas interface.
In Figure \ref{figure:13}, we can check that, in document ID01, both coders agreed to codify two quotations (of different length) with the second and third codes of P07.

\subsubsection{The $\alphabin$ coefficient}
\label{sec:atlas-alphabin}

\begin{figure}[!h]
\centering
\includegraphics [width=12.5cm, height=4cm]{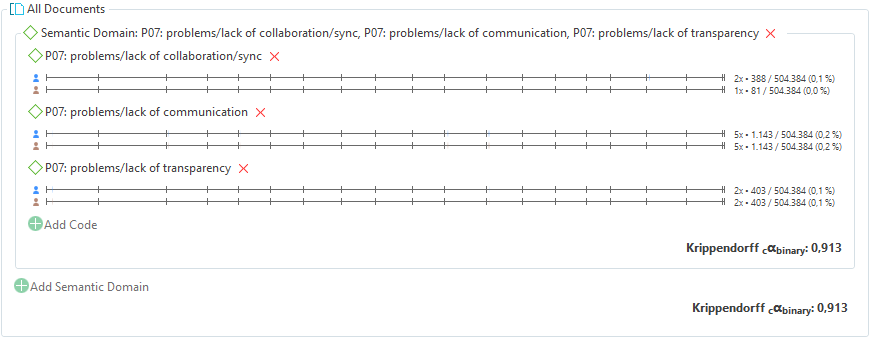}
\caption{Computation of the $\alphabin$ coefficient}
\label{figure:14}
\end{figure}

In order to compute this coefficient, click on the \texttt{Agreement Measure} button and select \texttt{Krippendorff's c-Alpha-binary} option. As it is shown in Figure \ref{figure:14}, the system returns two values. The first one is the $\alphabin$ coefficient per semantic domain (P07 in this case, with $\alphabin^{\textrm{P07}} = 0.913$) an another global coefficient of the domains as a whole that corresponds to what we called $\globalalphabin$ as described in Sections \ref{sec:globalalphabin} and \ref{sec:alphabin}. Since we selected a single semantic domain (P07), both values of $\alphabin$ and $\globalalphabin$ agree.


In the case shown in Figure \ref{figure:14}, the value of the coefficient is high ($\alphabin^{\textrm{P07}} = 0.913 > 0.8$) which can be interpreted as an evidence that the domain P07 is clearly stated, its boundaries are well-defined and, thus, the decision of applying it or not is near to be deterministic. However, observe that this does not measure the degree of agreement in the application of the different codes within the domain P07. It might occur that the boundaries of the domain P07 are clearly defined but the inner codes are not well chosen. This is not a task of the $\alphabin$, but of the $\cualpha$ coefficient. 

\begin{figure}[!h]
\centering
\includegraphics [width=0.65\textwidth]{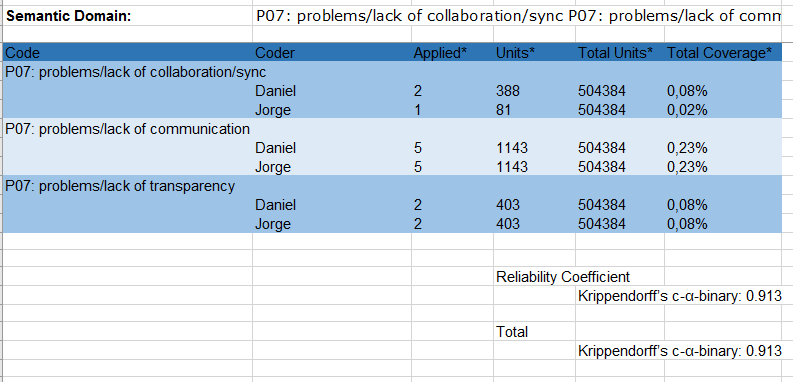}
\caption{Detailed coding information}
\label{figure:16}
\end{figure}

In order to illustrate how Atlas performed the previous computation, let us calculate $\alphabin$ by hand. For this purpose, we export the information provided by Atlas about the coding process. In order to do so, we click on the \texttt{Excel Export} button at the \texttt{Intercoder Agreement} panel.
In Figure \ref{figure:16} we show the part of the exported information that is relevant for our analysis. As we can see, there are two coders (Jorge and Daniel) and three codes. The meaning of each column is as follows:
\begin{itemize}
    \item Applied*: Number of times the code has been applied.
    \item Units*: Number of units to which the code has been applied.
    \item Total Units*: Total number of units across all selected documents, voted or not.
    \item Total Coverage*: Percentage of coverage in the selected documents
\end{itemize}

The length of the quotation (what is called units in Atlas) is expressed in number of characters. From this information, we see that coder Daniel voted 388 units (characters) with the first code of the domain (\textit{problems/lack of collaboration/sync}) while coder Jorge only voted 81 units with that code. For the others codes, both coders apply them to 1143 and 403 units, respectively. Indeed, as we will check later, the quotations that Jorge chose for applying P07 are actually a subset of the ones chosen by Daniel. Hence, Daniel and Jorge achieved perfect agreement when applying the second and third codes of P07 while Jorge only considered eligible 81 units for the first code of the 388 chosen by Daniel.

From these data, we can construct the observed coincidence matrix, shown in Table \ref{tab:coincidences-alphabin-P07}, as explained in Section \ref{sec:background-krip-alpha} (see also Section \ref{sec:universal-alpha}). Recall from Section \ref{sec:alphabin} that a label $1$ means that the coder voted the quotation with a code of the semantic domain (P07 in this case) and the label $0$ means that no code of the domain was applied. 

\begin{table}[h]
    \centering
    \begin{tabular}{|c|c|c|c|c|}
        \hline &   & \multicolumn{2}{c|}{Coder 2 (Jorge)} & \\\hline
         &    & $1$ & $0$ & \\\hline
        \multirow{2}{*}{Coder 1 (Daniel)} & $1$ & $o_{1,1} = 3254$ & $o_{1,2} =307$ & $t_1 = 3561$ \\\cline{2-5}
         & $0$ & $o_{1,2} =307$ & $o_{2,2} =1004900$ & $t_2 = 1005207$ \\\hline
         &  & $t_1 =3561$ & $t_2=1005207$ & $t = 1008768$\\\hline
    \end{tabular}
    \caption{Observed coincidences matrix for $\alphabin^{\textrm{P07}}$.}
    \label{tab:coincidences-alphabin-P07}
\vspace{-0.3cm}
\end{table}

This matrix is computed as follows. The number of units to which the coders assigned any code from domain P07 is $81 + 1143 + 403=1627$ in the case of Jorge and $388 + 1143 + 403 = 1934$ for Daniel. Since the choices of Jorge are a subset of the ones of Daniel, we get that they agreed in $1627 = \min(1627, 1934)$ units. Recall that $o_{1,1}$ counts ordered pairs of votes, so we need to double the contribution to get $o_{1,1} = 2 \cdot 1627 = 3254$. On the other hand, Jorge did not apply any code of P07 to $504384 - 1627 = 502757$ units, while Daniel did not apply them to $504384 - 1934 = 502450$, which means that they agreed on not to chose P07 in $502450 = \min(502757, 502450)$ units. Doubling the contribution, we get $o_{2,2} = 1004900$. Finally, for the disagreements we find that Daniel applied a code from P07 to 307 units that Jorge did not select, so we get that $o_{1,2} = o_{2,1} = 307$. Observe that we do not have to double this value, since there is already an implicit order in this votes (Daniel voted $1$ and Jorge voted $0$). From these data, it is straightforward to compute the aggregated quantities $t_1 = o_{1,1} + o_{1,2} = 3561$, $t_2 = o_{1,2} + o_{2,2} = 1005207$ and $t = t_1 + t_2 = 1008768$.

In the same vein, we can construct the matrix of expected coincidences, as explained in Section \ref{sec:background-krip-alpha} (see also \ref{sec:universal-alpha}). The value of the expected disagreements are
$$
    e_{1,2} = e_{2,1} = \frac{t_1 t_2}{t-1}= \frac{3561 \cdot 1005207}{1008767} = 3548.43.
$$
Analogously, we can compute $e_{1,1}$ and $e_{2,2}$. However, they are not actually needed for computing the $\alpha$ coefficient, so we will skip them. With these calculations, we finally get that
$$
D_0 = o_{1,2} = 307, \quad D_e = e_{1,2} = 3548.43, \quad
    \alphabin^{\textrm{P07}} = 1- \frac{D_o}{D_e} = 1-\frac{307}{3548.43} = 0.913.
$$

We want to notice again that the previous calculation is correct because Jorge voted with a code of P07 a subset of the quotations that Daniel selected for domain P07. We can check this claim using Atlas. For that purpose, click on \texttt{Show Documents Details} and review how the codes were assigned per document. In the case considered here, Table \ref{tab:codification-P07} shows an excerpt of the displayed information. To shorten the notation, the first code of P07 is denoted by 7a, the second one by 7b, and the third one by 7c. In this table, we see that all the voted elements coincide except the last 307 units corresponding to document ID17, that Daniel codified and Jorge did not.

\small
\begin{table}[!h]
    \centering
    \begin{tabular}{|c|c|c|}
        \hline \textbf{Document ID} & \textbf{Coder 1 (Daniel)} & \textbf{Coder 2 (Jorge)} \\\hline
        ID01 & 7b 1x112 \& 7c 1x306  & 7b 1x112 \& 7c 1x306 \\\hline
        ID03 & 7b 1x185 & 7b 1x185 \\\hline
        ID05 & 7b 1x159 & 7b 1x159 \\\hline
        ID10 & 7a 1x81 & 7a 1x81 \\\hline
        ID11 & 7b 1x314 & 7b 1x314 \\\hline
ID12 & 7b 1x373 \& 7c 1x 97  & 7b 1x373 \&
7c 1x 97  \\\hline
ID17 & 7a 1x307 & \\\hline
    \end{tabular}
    \caption{Codified units, per document, for the semantic domain P07.}
    \label{tab:codification-P07}
\vspace{-0.3cm}
\end{table}

\normalsize

\begin{figure}[!h]
\centering
\includegraphics [width=12cm, height=3.5cm]{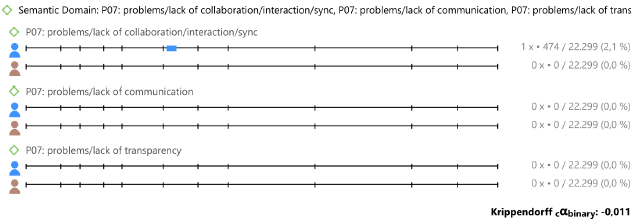}
\caption{Computation of the $\alphabin$ coefficient for the domain P07 in the second round}
\label{figure:18}
\end{figure}

Another example of calculation of this coefficient is shown in Figure \ref{figure:18}. It refers to the computation of the same coefficient, $\alphabin^{\textrm{P07}}$, but in the second round of the coding process (see Section \ref{sec:case-study-data-analysis}), where 11 documents were analyzed. We focus on this case because it shows a cumbersome effect of the $\alpha$ coefficient.

As shown in the figure, we get a value of $\alphabin^{\textrm{P07}} = -0.011$. This is an extremely small value, that might even point out to a deliberate disagreement between the coders. However, this is not happening here, but an undesirable statistical effect that fakes the result. The point is that, as shown in Figure \ref{figure:18}, in round 2 there is only one evaluation from one of the coders that assigns this semantic domain, in contrast with the 17 evaluations obtained in round 1. For this reason, there are not enough data for evaluating this domain in round 2 and, thus, this result can be attributed to statistical outliers.

The researcher interested in using Atlas for qualitative research should stay alert to this annoying phenomenon. When Atlas considers that there are not enough statistical evidences (p-value $> 0.05$) or the number of coded quotations is very small, these anomalous values can be obtained or even the text $\texttt{N/A}^*$ (Not Available). In this case, thanks to the few received codings, the reliability may be assessed by hand.

\subsubsection{The $\globalalphabin$ coefficient}

As we mentioned in Section \ref{sec:globalalphabin}, the $\globalalphabin$ coefficient allows researchers to measure the degree of agreement that the coders reached when distinguishing relevant and irrelevant matter. In this way, $\globalalphabin$ is only useful if each coder chops the matter by him/herself to select the relevant information to code. On the other hand, if the codebook designer pre-defines the quotations to be evaluated, this coefficient is no longer useful since it always attains the value $\globalalphabin = 1$. This later situation is the case in our running example. 

Recall that, as we mentioned in Section \ref{sec:atlas-alphabin}, the $\globalalphabin$ coefficient is automatically computed when we select the \texttt{c-Alpha-binary} option in Atlas. It is displayed at the bottom of the results, below all the $\alphabin$ coefficients, as a summary of the reliability in the semantic domains.

Nevertheless, as can be checked in Figure \ref{figure:14}, in the calculation performed in the previous Section \ref{sec:atlas-alphabin} we got a value $\globalalphabin = 0.931 < 1$, which apparently contradicts the aforementioned fact that with pre-defined quotations always perfect agreement is achieved. Recall that this coefficients evaluates the agreement of the coders when trying to discern which parts of the matter are relevant (quotations) and which ones are not. In other words, this coefficient distinguish coded matter (with any code) and non-coded matter. If we would introduce in Atlas all the domains, all the pre-defined quotations will have, at least, a code assigned, and the rest of the matter will not receive any code. In this way, we would get $\globalalphabin = 1$, since there exists perfect agreement between the relevant matter (the quotations) and the irrelevant matter. 

The key point here is that, in the calculation of Section \ref{sec:atlas-alphabin} we do not introduced in Atlas all the semantic domains, but only P07. In this way, the $\globalalphabin$ coefficient were not computed over the whole corpus. In other words, since we only added to the ICA tool the codes belonging to P07, the Atlas' analyzer considered that these are all the codes that exist in the codebook, and anything that did not receive a code of P07 is seen as irrelevant matter. In this way, the quotation (made of 307 units) that Daniel chose for applying the code 7a and Jorge did not is considered by the ICA tool as a quotation that Daniel saw as relevant matter and Jorge as irrelevant matter, decreasing the value of $\globalalphabin$.

\begin{figure}[!h]
\centering
\includegraphics [width=8cm, height=4.5cm]{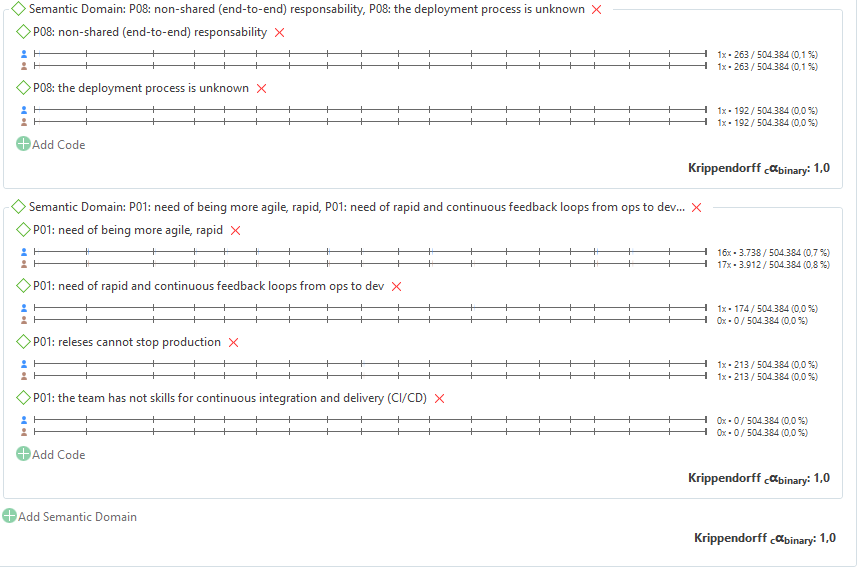}
\caption{Computation of the $\globalalphabin$ coefficient with domains P01 and P08}
\label{figure:20}
\end{figure}

If we conduct the analysis over all the codes of the codebooks it turns out that Jorge applied to this quotation a code from a different semantic domain so these 307 units are actually relevant matter, restoring the expected value $\globalalphabin = 1$. The same result is obtained if we select a subset of domains in which the coders used codes of the subset on the same quotations (even if they did not agree on the particular codes), as shown in Figure \ref{figure:20} for domains P01 and P08.

For this reason, it is crucial to evaluate the global $\alpha$ coefficients (namely, $\globalalphabin$ and $\Cualpha$) only when all the codes have been added to the ICA tool. Otherwise, the result might be wrong and may lead to confusions.

\subsubsection{The $\cualpha$ and $\Cualpha$ coefficients}

In some sense, the binary coefficients $\alphabin$ and $\globalalphabin$ evaluate whether to apply a particular semantic domain or not is well-defined in the codebook. They are binary measures, in the sense that they test a binary question: to apply some domain or not.

On the other hand, in this section we will consider the $\cualpha$ and $\Cualpha$ coefficients that, roughly speaking, zoom in to measure the limits of definition of the semantic domains themselves and of the codes within them. Recall from Section \ref{sec:cu-alpha} that the $\cualpha$ coefficient is computed per semantic domain. Hence, fixed a domain $S$, it evaluates the amount of agreement that the coders reached when choosing to apply some code of $S$ or other. It is, therefore, a measure of the reliability in the application of the codes withing $S$, not of the domains themselves. Analogously, as explained in Section \ref{sec:Cu-alpha}, $\Cualpha$ is a global measure that allows us to assess the limits of definitions of the semantic domains. In other words, it measures the goodness of the partition of the codebook into semantic domains, independently of the chosen code

These two later coefficients can be easily computed with Atlas. For that purpose, click on the option \texttt{Cu-Alpha/cu-Alpha} in the ICA pannel, under the \texttt{Agreement Measure} button. In Table \ref{tab:table-cualpha}, we show the obtained results of $\cualpha$ for each of 10 semantic domains of the running example considered in this section in the first round.

\small
\begin{table}[h]
    \begin{center}
    \begin{tabular}{|c c || c c|}
    \hline
        \textbf{Semantic domain} & $\cualpha$ & \textbf{Semantic domain} & $\cualpha$  \\
    \hline
         P01 & 0.705 &
    
         P02 & 1.0 \\
    
         P03 & 0.962  &
    
         P04 & 1.0 (N/A$\null^*$)\\
    
         P05 & 1.0 &
    
         P06 & 0.739 \\
    
         P07 & 1.0 &
    
         P08 & 1.0 \\
    
         P09 & 1.0 &
    
         P10 & 0.563 \\
    \hline
    \end{tabular}
    \caption{Values of $\cualpha$ per semantic domain}   
    \label{tab:table-cualpha}
    \end{center}
\vspace{-0.3cm}
\end{table}
\normalsize

\begin{figure}[!h]
\centering
\includegraphics [width=0.9\textwidth]{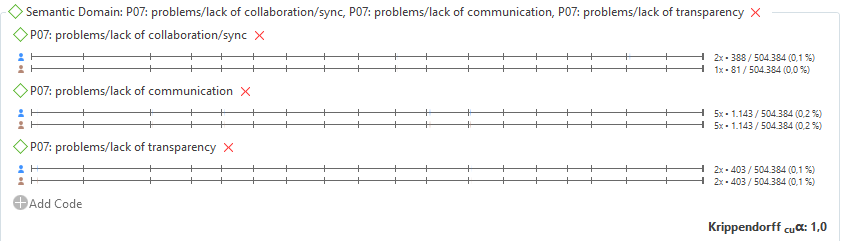}
\caption{Computation of the $\cualpha$ coefficient for the domain P07}
\label{figure:P07}
\end{figure}

Observe that $\cualpha$ attained its maximum value, $\cualpha = 1$, over the domains P02, P05, P07, P08, and P09. This may seem counterintuitive at a first sight since, as can be checked in Figure \ref{figure:P07} and as we mentioned in Section \ref{sec:atlas-alphabin}, in P07 there is no perfect agreement. Indeed, as we know the code of P07 ``\textit{problems/lack of collaboration/sync}'' was chosen by Daniel for a quotation that Jorge skipped. This is strongly related with Remarks \ref{ref:not-voted-item} and \ref{rmk:computation-alpha-non-voted} since recall that, fixed a domain $S$, in the observed coincidences matrix only quotations were voted with codes of $S$ by at least two different coders count. Otherwise, these quotations do not contribute with a pair of disagreements so, through the eyes of $\cualpha$, they do not compromise the reliability.

This fact is precisely what is taking place in this case. The quotation voted by Daniel with a code of P07 and not by Jorge does not appear in the observed coincidences matrix for $\cualpha^{\textrm{P07}}$, neither as an agreement or as a disagreement. This allows $\cualpha^{\textrm{P07}} = 1$ even though there is no perfect agreement in the evaluation. This might seem awkward, but it actually makes sense since this disagreement was already detected via $\alphabin^{\textrm{P07}} < 1$, so decreasing also $\cualpha^{\textrm{P07}}$ would count it twice. The same scenario occurs in domains P02 and P09.



\begin{figure}[!h]
\centering
\includegraphics [width=0.9\textwidth]{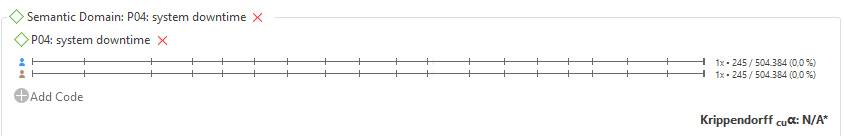}
\caption{Computation of the $\cualpha$ coefficient for the domain P04}
\label{figure:P04}
\end{figure}

Special mention deserves the semantic domain P04 (Figure \ref{figure:P04}), that was evaluated as $\texttt{N/A}^*$. Here, visual inspection shows that both coders do actually codify the same quotation with the same code. However, since only a single quotation was judged with a code of this domain, Atlas considers that there is not enough variability for providing statistical confidence (i.e.\ the p-value is above 0.05) to draw reliable conclusions. As we mentioned at the end of Section \ref{sec:atlas-alphabin}, a manual verification of the agreement is required in this case in order to assess the reliability.

\begin{figure}[!h]
\centering
\includegraphics [width=0.9\textwidth]{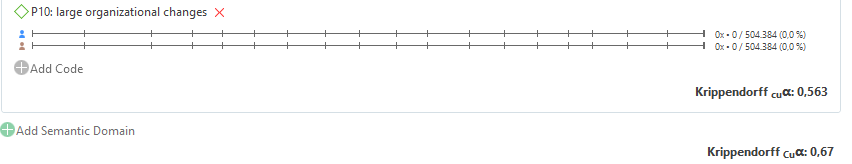}
\caption{Computation of $\Cualpha$ at the end of the semantic domains}
\label{figure:P10}
\end{figure}

Finally, as it can be checked in Figure \ref{figure:P10}, the $\Cualpha$ coefficient reached a value of $0.67$, which is slightly above the lower threshold of applicability of $0.667$. This suggests that the limits of definition of the semantic domains are a bit diffuse and can be improved. This problem was addressed in the second version of the codebook, in which a better definition of the domains allowed us to increase $\Cualpha$ to $0.905$, which is a sound evidence of reliability.

\section{Conclusions}
\label{sec:conclusions}

Along this tutorial, we have applied a set of statistical measures proposed in the literature for evaluating Inter-Coder Agreement in thematic analysis. Among them, we have payed special attention to Krippendorff's $\alpha$ coefficients as the most appropriate and best behaved for qualitative analysis.

After a formal introduction to these coefficients, we have presented a theoretical framework in which we provide a common formulation for all of them in terms of a universal $\alpha$ coefficient. This analysis provides a clearer and more precise interpretation of four of the most important variants of the $\alpha$ coefficients: the binary $\alpha$ coefficient ($\alphabin$), the global binary $\alpha$ coefficient ($\globalalphabin$), the cu coefficient ($\cualpha$) and the Cu coefficient ($\Cualpha$). This redefinition is particularly well suited for its use in case studies, interview surveys, and grounded theory, with a view towards providing sound reliability of the coding.

From an exploratory study about the adoption of the DevOps culture in software companies, and using the Atlas.ti software as a tool, in this paper we have presented a tutorial about how to apply these coefficients to software engineering research to improve the reliability of the drawn conclusions. With this idea in mind, we describe how to compute these coefficients using Atlas, and how to interpret them, leading to interpretations that complement the ones provided by the Atlas' user manual itself.

Furthermore, the interpretation provided in this paper has allowed us to detect some bizarre behaviors of the Krippendorff's $\alpha$ coefficients that are not described in the Atlas' manuals and that may mislead to researchers who are no familiar with the $\alpha$ measures. To shed light to these unexpected results, to justify why do they appear and how to interpret them have been guiding lines of this work. In particular, we addressed situations in which the insufficient statistical variability of the coding prevents Atlas to emit any measure of the attained agreement. We also explained paradoxical results in which very small deviations from the agreement lead to very bad measures of the $\alpha$ coefficient and we clarified why the $\cualpha$ may be maximum even though there is no perfect agreement and how to detect it through $\alphabin$.

Most of the qualitative works in software engineering suffer the lack of formal measures of the reliability of the drawn conclusions. This is a very dangerous threat that must be addressed to establish sound, well-posed and long-lasting knowledge. Otherwise, if the data are not reliable, the drawn conclusions are not trustworthy. In this direction, we expect that this tutorial will help researchers in qualitative analysis, in general, and in empirical software engineering, in particular, to incorporate these techniques to their investigations, aiming to improve the quality and reliability of the research.

\bibliography{references.bib}{}
\bibliographystyle{abbrv}

\end{document}